\documentclass[reprint, apm]{revtex4-2}
\usepackage{mathrsfs}
\usepackage{svg}
\usepackage{graphicx}
\usepackage{dcolumn}
\usepackage{bm}
\usepackage{amsmath,amssymb}
\usepackage{array}
\usepackage{booktabs} 
\usepackage{multirow}
\usepackage{subfigure}
\usepackage{natbib}
\usepackage{caption}
\usepackage{ragged2e}
\usepackage{subcaption} 

\begin{document}
\title{A Machine Learning Approach Capturing Hidden Parameters in Autonomous Thin-Film Deposition}

\author{Yuanlong Zheng$^{1,2}$}
\author{Connor Blake$^{2}$}
\author{Layla Mravac$^{2}$}
\author{Fengxue Zhang$^{3}$}
\author{Yuxin Chen$^{3}$}
\author{Shuolong Yang$^{2}$}

\affiliation{$^{1}$Department of Physics, University of Chicago, Chicago, IL 60637, USA}
\affiliation{$^{2}$Pritzker School of Molecular Engineering, University of Chicago, Chicago, IL 60637, USA}
\affiliation{$^{3}$Department of Computer Science, University of Chicago, Chicago, IL 60637, USA}

\date{\today}


\begin{abstract}
\textbf{Abstract}

The integration of machine learning and robotics into thin film deposition is transforming material discovery and optimization. However, challenges remain in achieving a fully autonomous cycle of deposition, characterization, and decision-making. Additionally, the inherent sensitivity of thin film growth to hidden parameters such as substrate conditions and chamber conditions can compromise the performance of machine learning models. In this work, we demonstrate a fully autonomous physical vapor deposition system that combines in-situ optical spectroscopy, a high-throughput robotic sample handling system, and Gaussian Process Regression models. By employing a calibration layer to account for hidden parameter variations and an active learning algorithm to optimize the exploration of the parameter space, the system fabricates silver thin films with optical reflected power ratios within 2.5\% of the target in an average of 2.3 attempts. This approach significantly reduces the time and labor required for thin film deposition, showcasing the potential of machine learning-driven automation in accelerating material development.
\end{abstract}

\maketitle

\section{Introduction}

In physical vapor deposition (PVD) of thin film materials, the traditional human-led process encompasses numerous cycles of selecting deposition parameters, performing deposition, characterizing film properties, and re-adjusting the parameters accordingly. The recent development of machine learning (ML)~\cite{Jordan2015}, coupled with advancements in robotics \cite{Garcia2007}, now potentially enables fully automating this process, liberating researchers from the repetitive cycles and accelerating the optimization of material properties \cite{Morgan2020}.

Several studies have sought to integrate ML with the PVD process. Typically, these approaches involved training ML models that map deposition parameters, such as substrate temperature, deposition rate, and flux ratio, to material properties, such as stoichiometry~\cite{Wakabayashi2023,Febba2023,Johnson2024Active}, electrical conductivity~\cite{ishiyama2024bayesian,shrivastava2024bayesian,Wakabayashi2019}, surface morphology~\cite{messecar2024quantum,shen2024machine}, crystallinity~\cite{kim2023machine,provence2020machine,Guevarra2022Materials,Ni2022Phase,Liang2022Application,Ament2021Autonomous}, and superconducting critical temperatures~\cite{Ohkubo2021}. The trained models are then used to predict material properties, with Bayesian optimization (BO) frequently employed to autonomously determine the deposition parameters for subsequent samples~\cite{Wakabayashi2019,Wakabayashi2022,Packwood2017Bayesian}. Such optimization has the potential to replace human decision-making, efficiently exploring the deposition parameter space to optimize the material properties~\cite{Shahriari2015}. 

However, ML models are known for their sensitivity to noise in the training data. In the meantime, the thin film deposition process can be prone to irreproducibility due to factors such as different initial substrate conditions and chamber environments~\cite{Febba2023,shrivastava2024bayesian,Ohkubo2021}. The inconsistency in the growth outcome may be treated by the model as noise which undermines the model's performance. These ``hidden parameters'' that lead to the irrepeatability of thin film properties are often difficult to capture and incorporate into the model. The field of ML-assisted thin film deposition currently lacks a systematic approach to effectively account for these hidden parameters.
\begin{figure*}[ht]
    \centering
    \includegraphics[width=\textwidth]{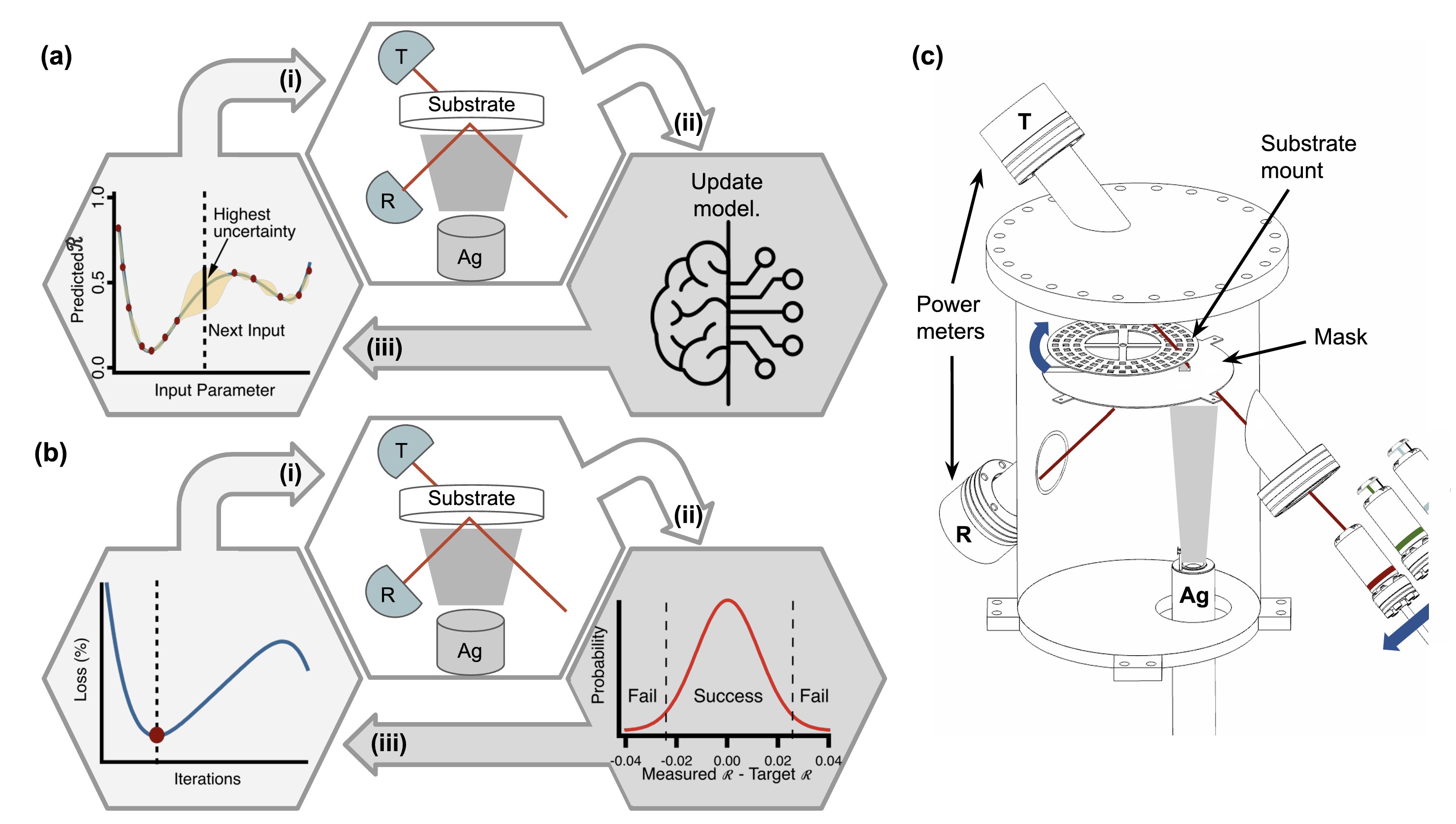} 
    \captionsetup{justification=raggedright,singlelinecheck=false}
    \caption{\justifying{A self-learning autonomous physical vapor deposition system for silver thin film growth. (a) The autonomous learning cycle incorporates (i) identification of the growth condition with the highest model uncertainty, (ii) sample growth, and (iii) updating the model with new data. (b) The autonomous testing cycle incorporating (i) prediction of the optimal growth condition that minimizes the loss function, (ii) sample growth, and (iii) assessment of success. (c) Schematic illustration of the autonomous deposition setup featuring robotic sample handling and in-situ optical characterization.}}
    \label{fig:1}
\end{figure*}

Beyond algorithmic challenges, the full potential of ML-assisted PVD hinges on the complete automation of hardware systems. PVD systems require high vacuum (HV) or ultra-high vacuum (UHV) environments, which present significant challenges for fully automating sample transfer and characterization processes. As such, most studies in the field of ML-assisted thin film deposition still rely on traditional manual handling of samples, which limits sample throughput and hinders the realization of the fully autonomous ML-assisted PVD~\cite{Wakabayashi2019,Ohkubo2021,kusne2020on}. Shimizu et al. demonstrated a system fully automating the deposition of Nb-doped $\mathrm{TiO}_2$ films and the minimization of its resistance~\cite{Shimizu2020}. However, the system requires complex multi-chamber systems with sophisticated transfer mechanisms, thereby increasing the complexity of the setup and limiting its scalability. Achieving a scalable method to fully automate PVD systems is a critical hurdle to unlocking the full promise of ML integration in thin film deposition.

\section{Methods and Results}

In this work, we address inefficiencies in current ML-driven thin film deposition by developing a fully automated system with high sample throughput and a mechanism to account for hidden parameter variations. Our system integrates a UHV chamber with a 72-slot robotic sample handling system optimized for maximizing throughput, in-situ optical characterization, and ML-powered prediction of characterization outputs and optimal growth conditions. To account for the varying hidden parameters, we introduce a calibration layer that provides critical information about the initial condition of each sample and the chamber environment. Furthermore, we employ an autonomous learning approach using Gaussian Process Regression (GPR) models to efficiently explore the multidimensional input space and accurately predict the target properties of the films. By integrating these techniques, we have engineered an ML-driven, closed-loop deposition system, thereby eliminating the need for human intervention at any stage of the growth cycle. We demonstrate the autonomous synthesis of silver thin films with optical properties within 2.5\% of the user-specified target in an average of 2.3 attempts, underscoring the potential of our ML-driven system for significantly accelerating the optimization of material properties.

\noindent \textbf{1. Introduction of the System}

To demonstrate the principles of ML-driven autonomous PVD, we seek to fabricate silver thin films with user-specified optical properties. The growth conditions of silver films such as the growth rate, substrate temperature, and film thickness~\cite{Zhao2009} impact the films' porosity~\cite{Savaloni2008}, grain size~\cite{Reddy2017}, and electron-phonon interactions~\cite{Choi2020}. All of these aspects are correlated with the real and imaginary parts of the optical constants for silver films. It is difficult to model all of these mechanisms using simple physics laws, which warrants an ML-driven material optimization.    

Our autonomous PVD system incorporates a shadow mask beneath a 72-slot sample handling system, ensuring that only one sample is exposed to the molecular beam at a time (Figure 1(c)). Silver (99.999\%, Thermo Fisher) is deposited onto double-side polished BK7 substrates using an effusion cell (MBE-Komponenten) at a base pressure of $< 5\times{10}^{-9}$ mbar and a deposition pressure of ${1\times10}^{-8}$ mbar. The reflectivity and absorptivity of the silver thin films are characterized using five $p$-polarized lasers with wavelengths of 443, 514, 689, 781, and 817 nm (Coherent StingRay). The lasers are mounted on a computer numerically controlled (CNC) linear rail (Figure 1(c)) pointing at the substrate with an incident angle of 45 degrees. The transmitted (\(P_t\)) and reflected power (\(P_r\)) are measured, and we define the reflected power ratio $\mathscr{R}$ and absorptivity $\mathscr{A}$:

$$\mathscr{R} = \frac{P_r}{P_r + P_t},\quad \mathscr{A} =  \frac{P_i - P_r - P_t}{P_i}$$

\noindent where $P_i$ denotes the incident power. For convenience, the reflected power ratio and the absorptivity at 443 nm are denoted as $\mathscr{R}_{443}$ and $\mathscr{A}_{443}$, respectively, and similarly for other wavelengths.

All hardware control and data acquisition are managed by MATLAB scripts which allow the system to grow and characterize up to 72 samples consecutively without human intervention. The optical characterization data are fed into the ML algorithm, which predicts the growth condition required for the sample to attain user-specified $\mathscr{R}$ values at targeted wavelengths. The process is divided into the autonomous learning and autonomous testing stages. In the learning stage, optical characterization data are used to train ML models, which then determine the next point to explore for further learning (Figure 1(a)). In the testing stage, the trained model guides the system to fabricate silver thin films with user-specified optical properties (Figure 1(b)).

\noindent \textbf{2. Calibration Layer}

For thin film deposition, slight differences in substrate and chamber conditions can significantly affect growth dynamics, especially for the first few atomic layers~\cite{PhysRevX.11.021054}. Parameters such as substrate surface roughness, and residual pressure of various elements in the chamber are unknown to us. Although some of these parameters can be measured with various difficulties, it is infeasible for one to exhaust all the factors in the deposition process. To quantify the effects of these hidden parameters on the sample properties, we first grow 24 samples at an effusion cell temperature ($T$) of 875°C and a deposition time ($t$) of 5000 seconds. After 5000 seconds, the $\mathscr{R}_{817}$ of the samples has a distribution with a standard deviation of 0.061 (Figure 2(c)). 
\begin{figure}[h]
    \centering
    \raisebox{14\height}{\llap{\textbf{(a)}\hspace{-2em}}}
    \raisebox{14\height}{\llap{\textbf{(b)}\hspace{-15em}}}
    \begin{minipage}[t]{0.48\linewidth} 
        \centering
        \includegraphics[width=\linewidth]{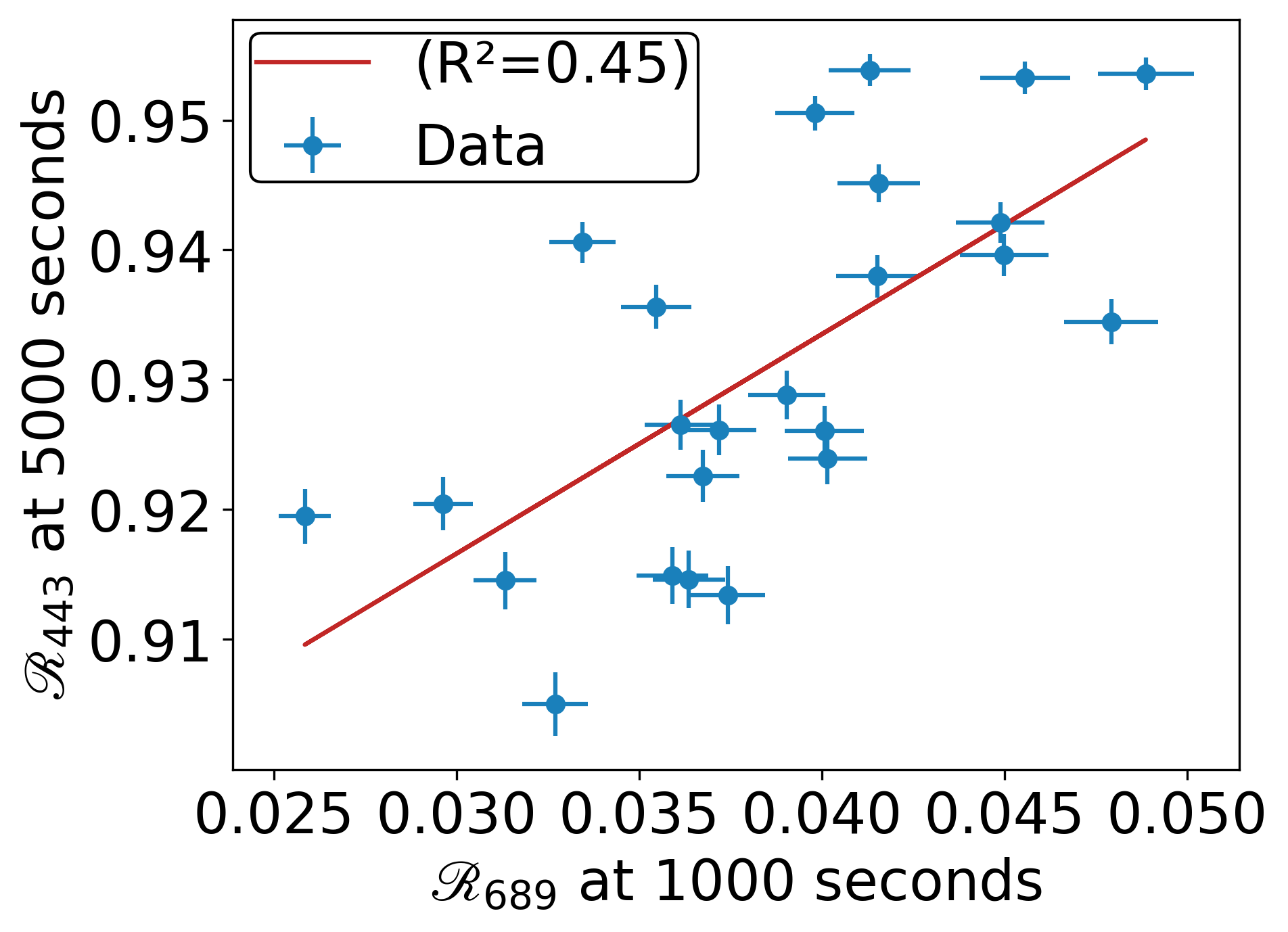}
        \label{fig:exampleA}
    \end{minipage}
    \hfill
    \begin{minipage}[t]{0.48\linewidth} 
        \centering
        \includegraphics[width=\linewidth]{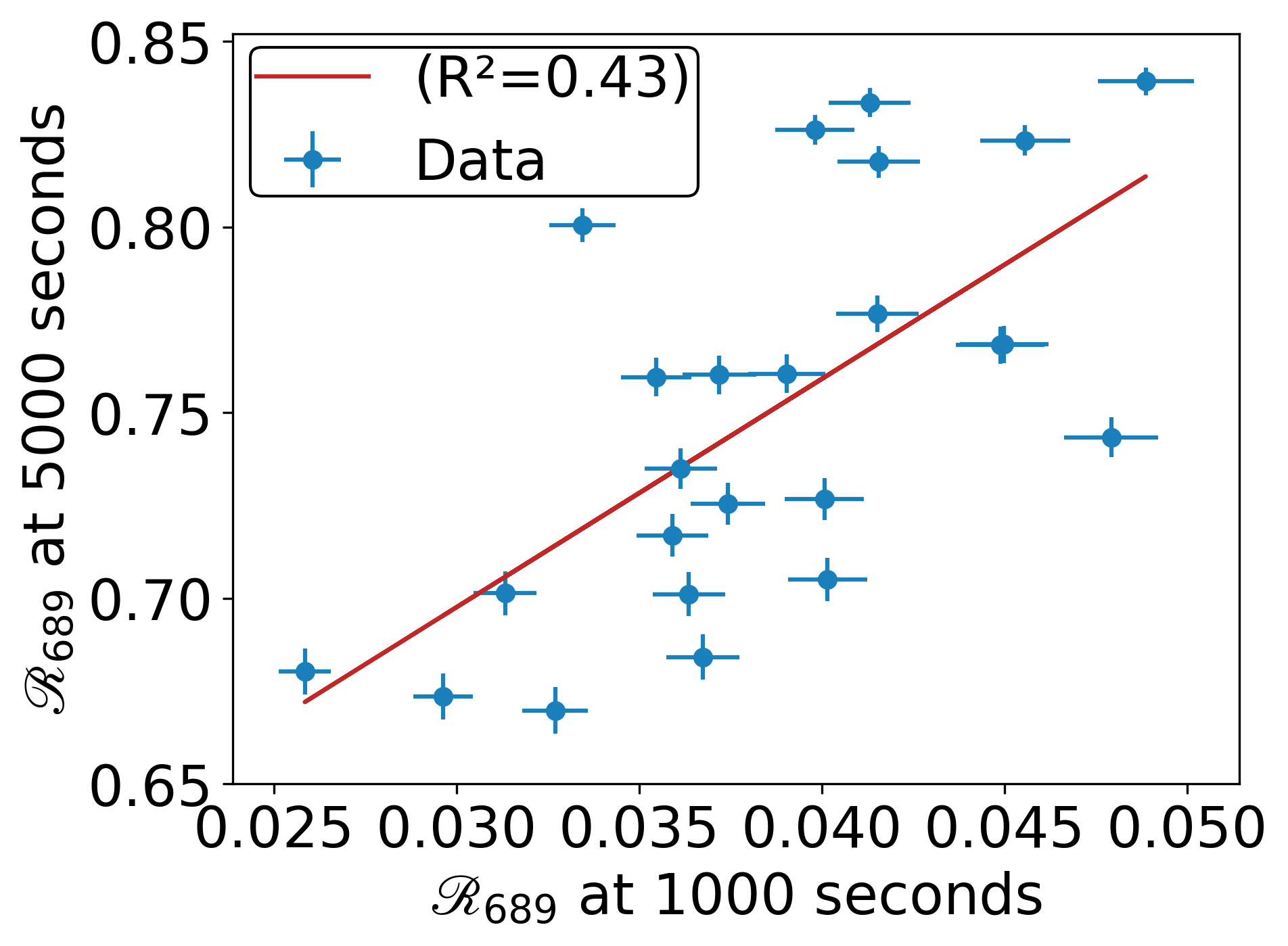}
        \label{figure2}
    \end{minipage}
    \vspace{-1em}
    
    \raisebox{14\height}{\llap{\textbf{(c)}\hspace{-2em}}}
    \raisebox{14\height}{\llap{\textbf{(d)}\hspace{-15em}}}
    \begin{minipage}[t]{0.48\linewidth} 
        \centering
        \includegraphics[width=\linewidth]{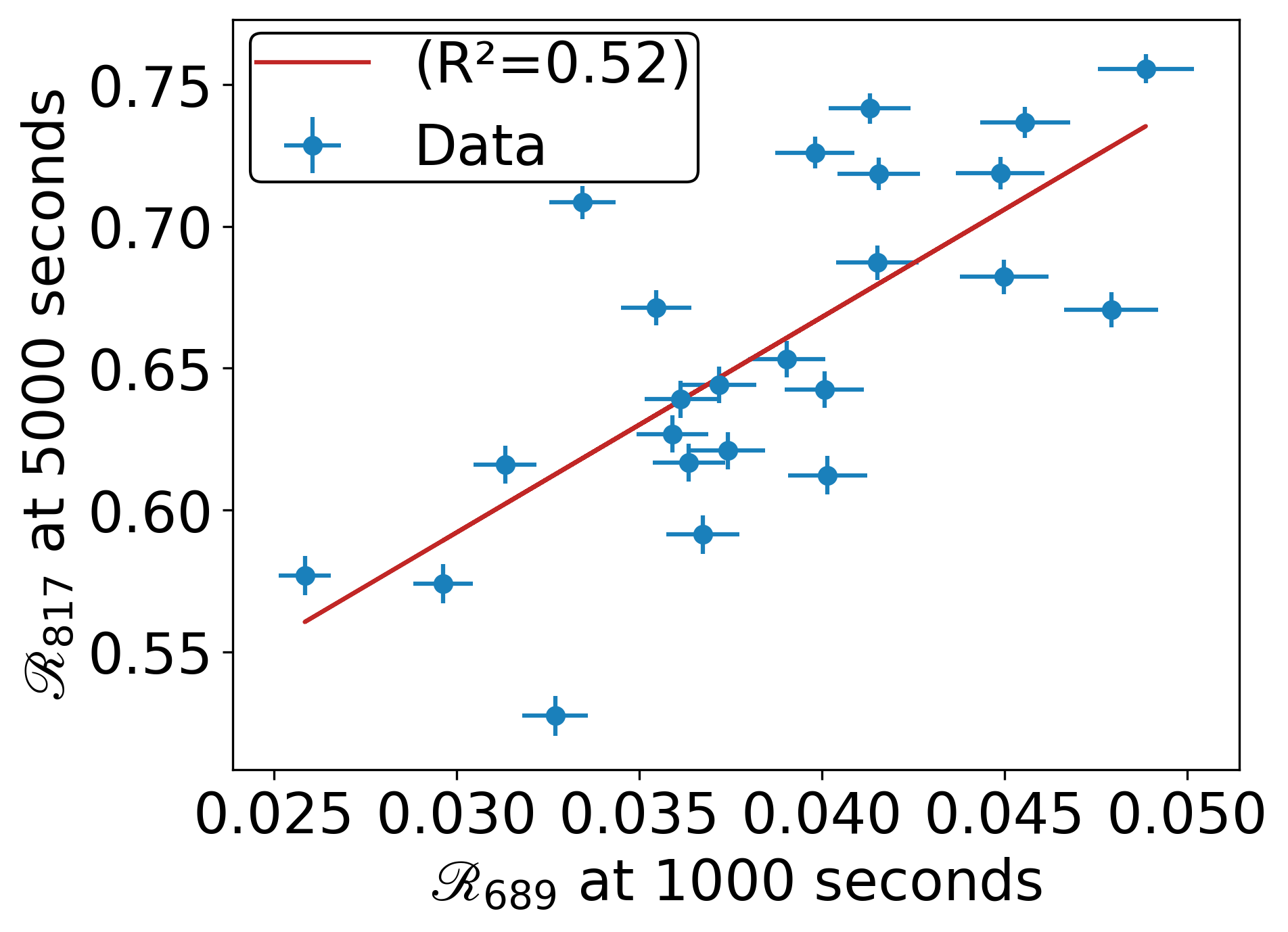}
        \label{fig:exampleA}
    \end{minipage}
    \hfill
    \begin{minipage}[t]{0.48\linewidth} 
        \centering
        \includegraphics[width=\linewidth]{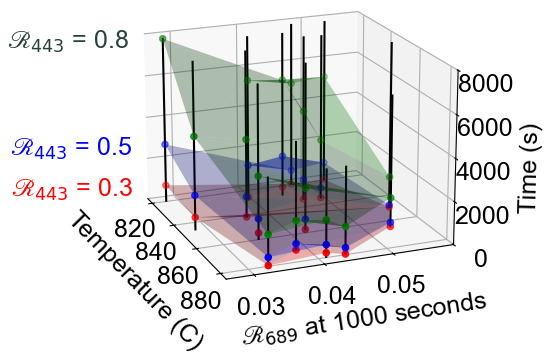}
        \label{figure2}
    \end{minipage}
    \captionsetup{justification=raggedright,singlelinecheck=false}
    \caption{\justifying{Calibration layer accounts for variations in optical properties of silver films. (a-c) A linear fit between $\mathscr{R}_{443}$, $\mathscr{R}_{689}$, $\mathscr{R}_{817}$ at 5000 seconds and $\mathscr{R}_{689}$ at 1000 seconds. The moderate R-squared values indicate that the variances of $\mathscr{R}$'s at later times can be partially captured by $\mathscr{R}$ at 1000 seconds. (d) Translucent planes of time required to reach specified $\mathscr{R}_{443}$ values for given effusion cell temperatures and $\mathscr{R}_c$'s. $\mathscr{R}_c$ is negatively correlated with the time required to reach a given $\mathscr{R}_{443}$. }}
\end{figure}

To account for this variability, we find that the $\mathscr{R}$'s of the 24 samples measured at $t=1000$ seconds can be used to quantify the fluctuations in the growth dynamics and its correlation with $\mathscr{R}$'s at later times. Here we plot $\mathscr{R}_{443}$, $\mathscr{R}_{689}$, $\mathscr{R}_{817}$ at 5000 seconds against their $\mathscr{R}_{689}$ at 1000 seconds and fit them with a simple linear function (Figure 2 (a-c)). The standard deviation of the $\mathscr{R}_{817}$ at 5000 seconds is 0.061. Meanwhile, its root-mean-square deviation (RMSD) from the linear fit is 0.032. This suggests that $\mathscr{R}_{689}$ at 1000 seconds is moderately correlated with the variation of $\mathscr{R}$'s at later times. Therefore, by initially growing a thin layer at \(T = 875^\circ\text{C}\) for 1000 seconds and measuring its $\mathscr{R}_{689}$, we partially captures the effect of the hidden parameters on the growth outcome. We define this layer as the \textit{calibration layer}, and its $\mathscr{R}_{689}$ is denoted as $\mathscr{R}_c$. We choose 689 nm as the wavelength for $\mathscr{R}_c$ due to its median position among the five wavelengths, thereby providing the most representative information. 

For all samples in the subsequent experiments, we first grow the calibration layer and record $\mathscr{R}_c$. The sample is then grown for time $t$ at temperature $T$. $\mathscr{R}$ and $\mathscr{A}$ are collected for all wavelengths throughout the growth (Figure 3). To maximize data acquisition efficiency, we divide the total deposition time into 98-second blocks. Within each block, the reflected and transmitted power at each wavelength are sequentially measured starting at the time points 0, 16, 32, 48, and 64 seconds (Table 1). The intervals between measurements are limited by the speed of the linear rail and the collection period of 5 seconds per wavelength. Note that $\mathscr{R}_c$, though a measured value, is treated as an input parameter of the model in the dataset. 

\begin{figure}[h]
    \centering
    \includegraphics[width=0.97\linewidth]{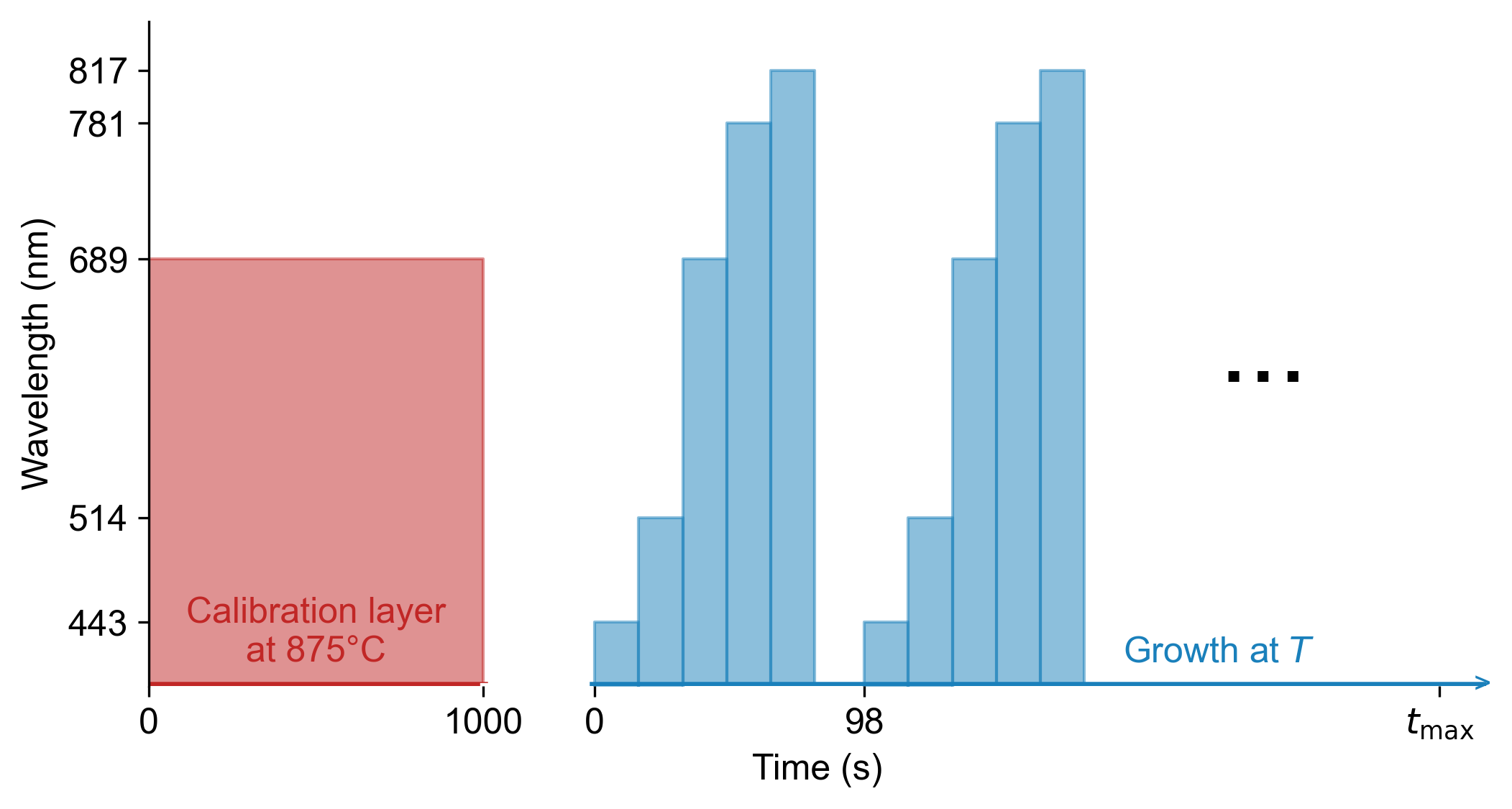}
    \captionsetup{justification=raggedright,singlelinecheck=false}
    \caption{\justifying{Illustration of the data acquisition process for each iteration in the autonomous learning stage: the calibration layer is first grown to determine the $\mathscr{R}_c$, followed by measuring optical properties at all 5 wavelengths in cycles.}}
    \label{fig:3}
\end{figure}

\begin{table}[h]
    \centering
    \resizebox{0.97\linewidth}{!}{ 
        \renewcommand{\arraystretch}{1.3} 
        \begin{tabular}{cccc@{\hskip 20pt}cc} 
        \toprule
        \multicolumn{4}{c}{\textbf{Model Inputs}} & \multicolumn{2}{c}{\textbf{Model Outputs}} \\ 
        \cmidrule(lr){1-4} \cmidrule(lr){5-6} 
        \textbf{Temperature (°C)} & \textbf{Time (s)} & \textbf{$\mathscr{R}_c$} & \textbf{Wavelength (nm)} & \multicolumn{1}{c}{$\mathscr{R}$} & \multicolumn{1}{c}{$\mathscr{A}$} \\ 
        \midrule
        872.5 & 0   & 0.038 & 443 & 0.356 & 0.422 \\
        872.5 & 16  & 0.038 & 514 & 0.158 & 0.194 \\
        872.5 & 32  & 0.038 & 689 & 0.044 & 0.114 \\
        872.5 & 48  & 0.038 & 781 & 0.039 & 0.048 \\
        872.5 & 64  & 0.038 & 817 & 0.034 & 0.014 \\
        872.5 & 98  & 0.038 & 443 & 0.394 & 0.442 \\
        \midrule
        872.5 & \dots & 0.038 & \dots & \dots & \dots \\
        872.5 & $t_{\text{max}}$ & 0.038 & 817 & 0.824 & 0.363 \\
        \bottomrule
        \end{tabular}
    }
    \captionsetup{justification=raggedright,singlelinecheck=false}
    \caption{\justifying{An exemplary data set taken at the effusion cell temperature $872.5$~\textcelsius, with model inputs and outputs.}}
    \label{tab:1}
\end{table}

\noindent \textbf{3. Autonomous Learning}

GPR models are a form of probabilistic supervised learning models derived from non-parametric statistics. GPR models are defined as a continuous set of distributions $f(\mathbf{x})$ where $f$ maps values $\mathbf{x} \in \mathbb{R}^n$ to the space of Gaussian probability density functions on $y$. For any two data points $(\mathbf{x}_1,y_1)$ and $(\mathbf{x}_2,y_2)$, the covariance between $y_1$ and $y_2$ is assumed to be purely a function of $\mathbf{x}_1$ and $\mathbf{x}_2$, $k(\mathbf{x}_1,\mathbf{x}_2)$.

A GPR model is fully characterized by its prior function \( g(\mathbf{x}) \), mean function \( m(\mathbf{x}) \equiv \mathbb{E}[f(\mathbf{x})] \), and covariance function \( k(\mathbf{x}_1, \mathbf{x}_2) \). The prior function \( g(\mathbf{x}) \) describes a "guess" for how the output space should appear in the absence of data. The mean function describes the expected \( y \) value for a given input \( \mathbf{x} \) based on the set of data points \(\{(\mathbf{x}_i, y_i)\}\). The covariance function, referred to as a kernel, describes how well one should be able to predict \( \Tilde{y} \) given \( \Tilde{\mathbf{x}} \) and \((\mathbf{x}, y)\). Here, we used a radial basis function (RBF) kernel.

\begin{equation}
    k_{\text{RBF}}(\mathbf{x}_1,\mathbf{x}_2) = \exp{\left(-\frac{1}{2}(\mathbf{x}_1-\mathbf{x}_2)^T \mathbf{\Theta^{-2}} (\mathbf{x}_1-\mathbf{x}_2)\right)}
\end{equation} chosen for its ability to act as a universal approximator \cite{Rasmussen2006Gaussian}. The length scale matrix $\mathbf{\Theta} = \text{diag}(\theta_1, \theta_2, \ldots, \theta_n)$ contains the length-scale parameters which are optimized during learning and describe the width of the Gaussian kernel used for prediction. 

Mapping the data and prior distribution to the predicted output (the posterior distribution) is known as conditioning and is computed via Bayes' rule. As such, the predicted posterior distribution at a point $\mathbf{\Tilde{x}}$ based on observed data $\{(\mathbf{x}_i,y_i)\}$ is given by a normal distribution:

\begin{equation}
    \label{eq:posterior_conditioning}
    f(\mathbf{\Tilde{x}}) \sim \mathcal{N}(k_i (K^{-1})_{ij} y_j, k(\mathbf{\Tilde{x}},\mathbf{\Tilde{x}}) - k_i (K^{-1})_{ij} k_j)
\end{equation}
where $k_i = k(\mathbf{\Tilde{x}},\mathbf{x}_i)$ and $K_{ij} = k(\mathbf{x}_i,\mathbf{x}_j)$ for symmetric kernels like RBF. Here we have used the prior $g(\mathbf{x}) = 0$ for simplicity. From here, this distribution will be written as $\text{GPR}_y(\mathbf{x}) = (\mu_y, \sigma_y)$ with $\mu_y$ and $\sigma_y$ referred to as the ``predicted mean'' and ``predicted uncertainty'' respectively. While this distribution appears to be computationally intractable for large datasets, the function $m(\mathbf{x})$ has a closed-form representation for the RBF kernel and is readily computed by libraries such as GPyTorch \cite{gardner2021gpytorchblackboxmatrixmatrixgaussian}.

We employ two GPR models for $\mathscr{R}$ and $\mathscr{A}$ with RBF kernels, both having input parameters $\mathbf{x} = (T, t, \lambda, \mathscr{R}_c)$, and outputs $(\mu_R, \sigma_R)$ and $(\mu_A, \sigma_A)$, respectively. Here, we define two modes of generating training data for models: predefined learning and autonomous learning. In predefined learning, the growth parameters of the training samples are set to divide the parameter space into equal increments. In autonomous learning, the model autonomously determines these growth parameters in a way that optimally learns the parameter space.

A series of predefined learning is used to initialize the model during which 9 samples are grown between 820 and 880°C with 7.5°C increments and for $t$ of 
\begin{equation}
t_{\text{max}}(T) = 3.22 \times 10^{14} \times e^{-0.0285T(^\circ C)} (\text{seconds})
\end{equation}
where $t_{\text{max}}$ is determined such that each sample reaches a thickness yielding an $\mathscr{R} > 0.8$ for all wavelengths. This ensures the generation of data over a wide range of $\mathscr{R}$ values for training the model, while also avoiding the unnecessary time spent growing films until $\mathscr{R}$ asymptotically approaches 1. The exponential relationship between $t_{\text{max}}$ and $T$ assumes that silver's vapor pressure, and consequently the growth rate, increases exponentially with $T$, ensuring that all samples at $t_{\text{max}}$ have approximately the same thickness. During the growth of each sample, $\mathscr{R}$ and $\mathscr{A}$ are repeatedly measured at each wavelength to obtain data between $t = 0$ and $t_{\text{max}}$ (Table 1).

The system proceeds to the autonomous learning stage. After each sample's calibration layer is grown and $\mathscr{R}_c$ is measured, the effusion cell temperature $T$ for the remainder of the growth is selected within the interval of [820, 880]~\textdegree C according to:  

\begin{equation}
    T_{\text{selected}} = \arg \max_{T} \left\{ \frac{ \sqrt{\sum_{t=0}^{t_{\text{max}}(T)} \sum_{\lambda} \sigma_{R}(T, t, \lambda, \mathscr{R}_c)^2}}{t_{\text{max}}(T)} \right\}
\end{equation}

\begin{figure}[h]

    \centering
    \raisebox{0\height}{\llap{\textbf{(a)}\hspace{12em}}}
    \includegraphics[width=\linewidth]{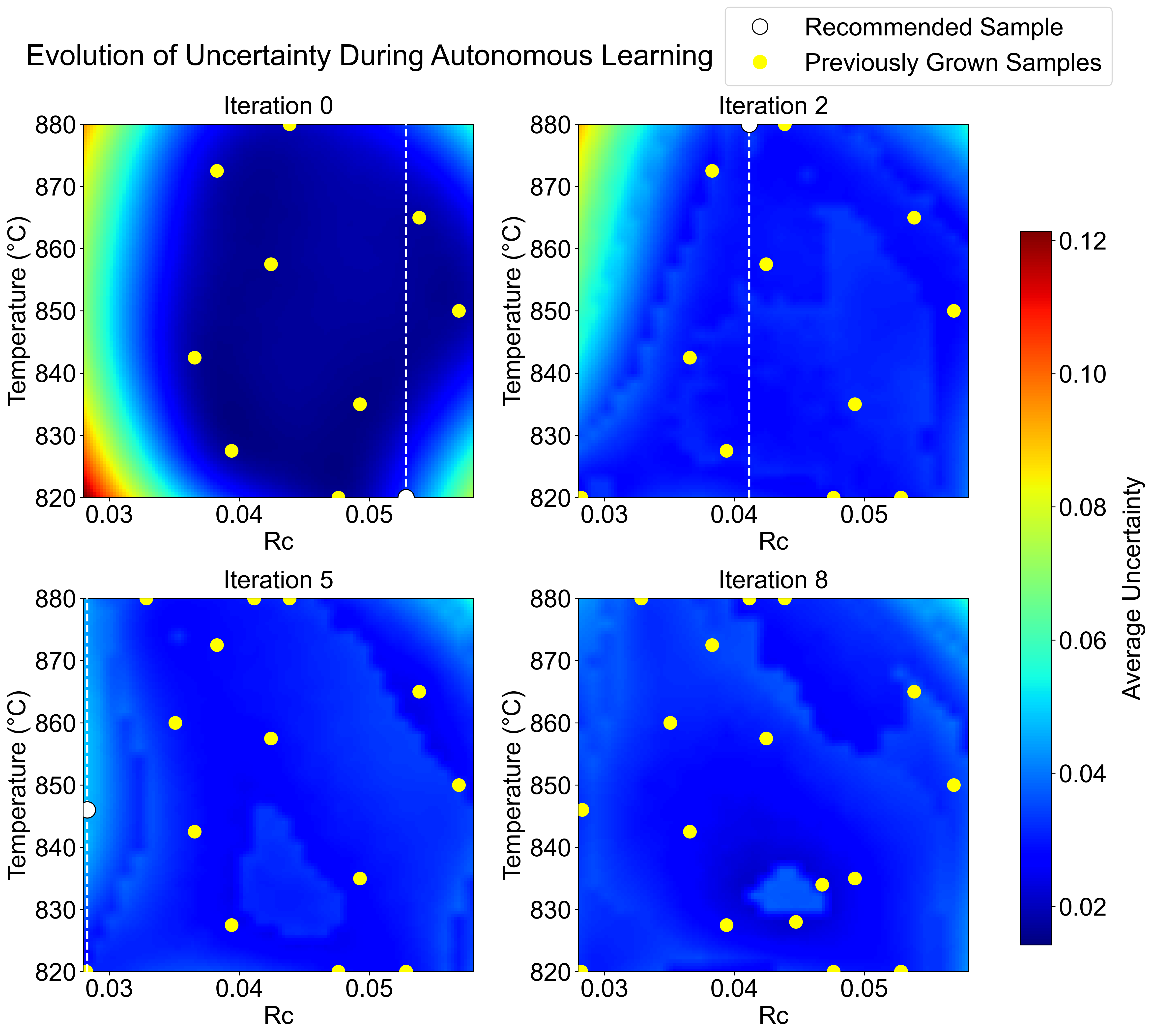}
    \label{fig:example}
    \raisebox{15\height}{\llap{\textbf{(b)}\hspace{-2em}}}
    \raisebox{15\height}{\llap{\textbf{(c)}\hspace{-15em}}}
    \begin{minipage}[t]{0.47\linewidth} 
        \centering
        \includegraphics[width=\linewidth]{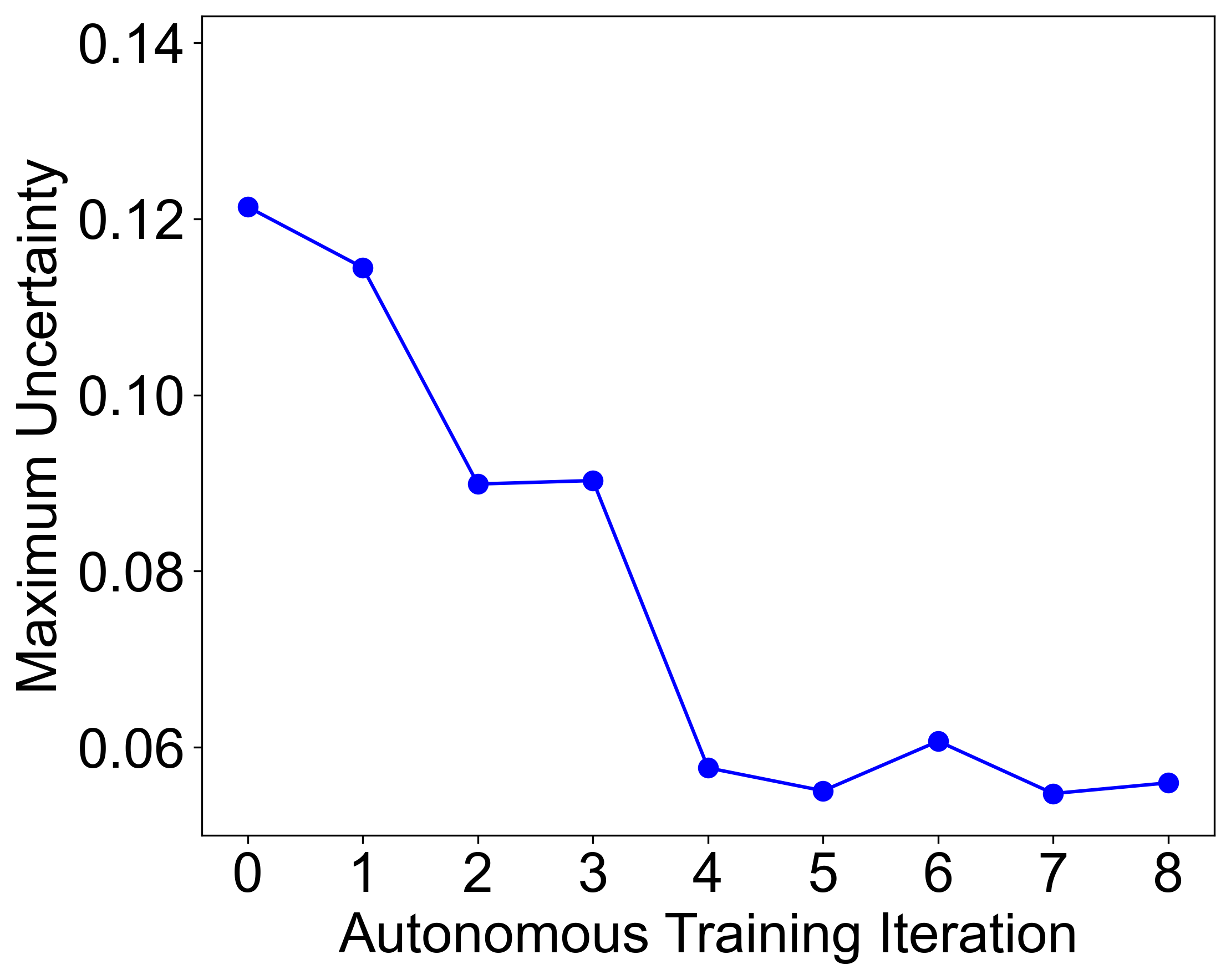}
        \label{fig:exampleA}
    \end{minipage}
    \hfill
    \begin{minipage}[t]{0.47\linewidth} 
        \centering
        \includegraphics[width=\linewidth]{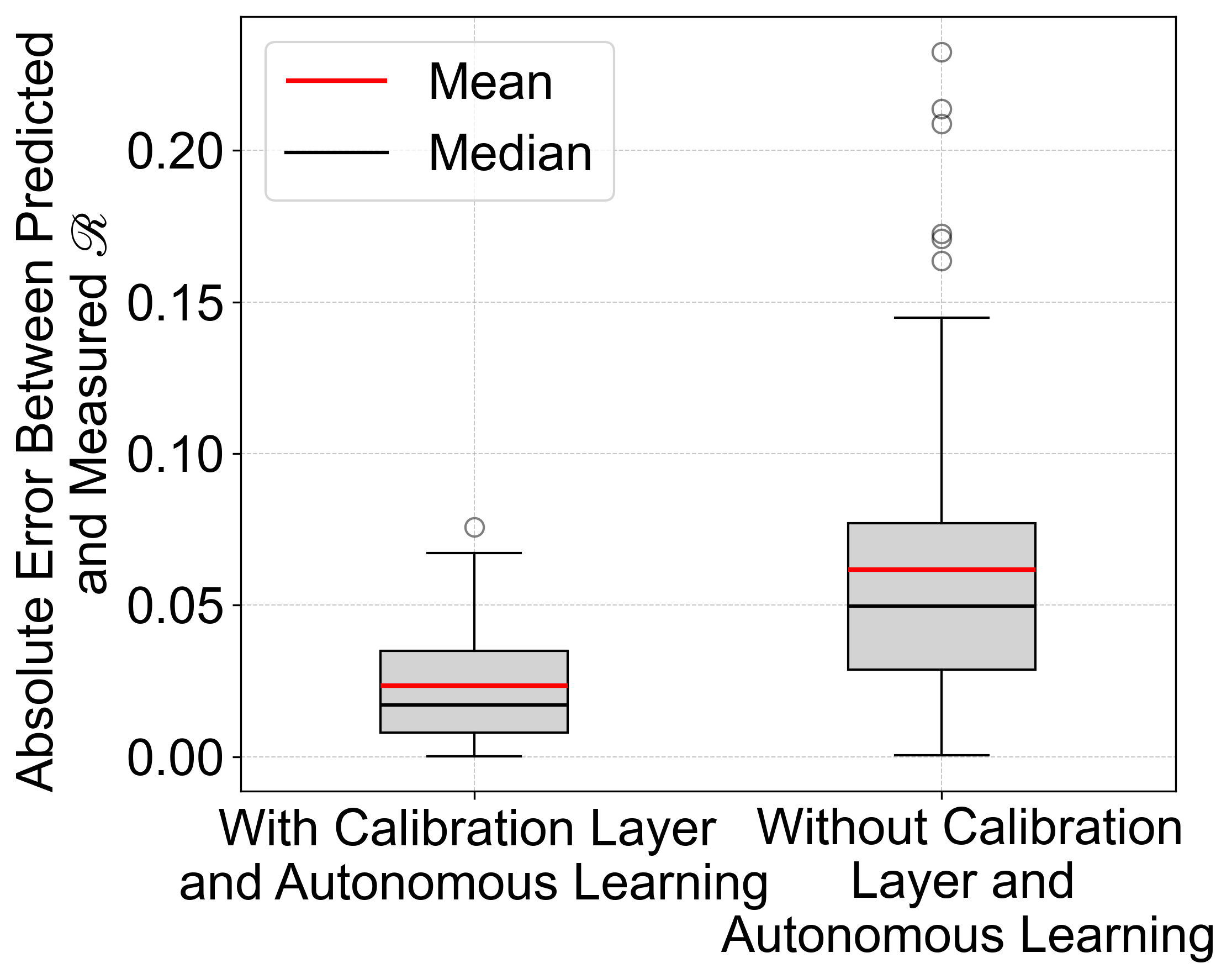}
        \label{fig:exampleB}
    \end{minipage}
    \captionsetup{justification=raggedright,singlelinecheck=false} 
    \caption{\justifying{Performance of autonomous learning in the fabrication of silver thin films. (a) Evolution of model predicted uncertainty, averaged over all $\mathscr{R}_{\lambda}$'s, during autonomous learning. (b) Convergence of the maximum uncertainty during autonomous learning. (c) Comparing the model prediction errors, defined as the differences between all measured and predicted $\mathscr{R}_{\lambda}$'s. The error distribution when adopting the calibration layer and autonomous learning is benchmarked against the case without these techniques.}}
\end{figure}
Here the $T$ with the highest prediction uncertainty for all relevant times and wavelengths is selected for the growth. The sample is then grown and measured at a series of times, as shown in Figure (3), until $t = t_{\text{max}}(T_{\text{grow}})$ specified by Eqn. (3). Such a complete deposition and characterization process, until the growth time reaches $t_{\text{max}} (T_{grow})$, is termed \textit{one iteration} (Figure 3). Note that the model only needs to choose $T$, among all input parameters ($T$, $t$, $\lambda$, $\mathscr{R}_c)$, because data are collected for each sample repeatedly over time and for all wavelengths, while $\mathscr{R}_c$ is a result of the current chamber and substrate initial condition and cannot be controlled by the algorithm. After each iteration of deposition and characterization is finished, the GPR model is updated with the new data. This process is repeated for 8 iterations until the maximum uncertainty in the parameter space converges (Figure 4 (b)). The total learning stage consists of 17 samples, including 9 from predefined learning and 8 from autonomous learning.

Figure 4(a) shows the evolution of the average predicted uncertainty for all $\mathscr{R}$'s during autonomous learning. After 8 iterations, the uncertainty over the entire parameter space becomes more uniformly distributed and has an average value of 0.032. The uncertainty does not further decrease below this level. The remaining uncertainty is likely due to the hidden parameters unaccounted by the calibration layer and measurement noises (e.g. uncertainty in laser power measurements). 

Unlike typical Bayesian optimizations aimed at optimizing a single output \cite{Greenhill2020}, our model is trained to acquire comprehensive information over the entire parameter space, enabling the system to respond to arbitrary $\mathscr{R}$ requests at given wavelengths.

\noindent \textbf{4. Autonomous Testing}

4.1	Single-wavelength $\mathscr{R}$ Targets

We select 5 random single-wavelength targets $\mathscr{R}_{\lambda}^{\text{target}}$, one for each wavelength of the lasers. Given a certain $\mathscr{R}_c$, there exist infinitely many $(T, t)$ that can achieve the requested $\mathscr{R}$ at the specified wavelength. The degeneracy is removed by also aiming for the minimum $\mathscr{A}_{\lambda}$. The loss function for each single-wavelength target is defined as 
\begin{flalign}
\mathscr{L}_{\lambda} &= (\mu_{\mathscr{R},\lambda} - \mathscr{R}_{\lambda}^{\text{target}})^2 + 4\sigma_{\mathscr{R},\lambda}^2 + \mu_{\mathscr{A},\lambda}^2 + 4\sigma_{\mathscr{A},\lambda}^2 & \nonumber \\
&\text{if } |\mu_{\mathscr{R},\lambda} - \mathscr{R}_{\lambda}^{\text{target}}| > 0.01: & \nonumber \\
&\quad \mathscr{L}_{\lambda} \mapsto \mathscr{L}_{\lambda} + 100 (\mu_{\mathscr{R},\lambda} -\mathscr{R}_{\lambda}^{\text{target}})^2 &
\end{flalign}

The uncertainties of predicted $\mathscr{R}$ and $\mathscr{A}$ are added to the loss function to penalize growth conditions with high uncertainties. The loss increases rapidly when the difference between the target and predicted $\mathscr{R}$ exceeds 0.01 to prioritize proximity to the target value of $\mathscr{R}$.

The loss function is minimized using the Adam optimizer \cite{kingma2017adammethodstochasticoptimization}, an adaptive learning rate optimization algorithm designed to handle sparse gradients on noisy problems. The set of growth parameters $(T, t)$ at the minimum of the loss function is used for deposition. After deposition, the sample's measured $\mathscr{R}_{\lambda}$ is compared to $\mathscr{R}_{\lambda}^{\text{target}}$. If $|\mathscr{R}_{\lambda} - \mathscr{R}_{\lambda}^{\text{target}}| < 0.025$, the growth is considered successful, and the system moves to the next target. If unsuccessful, the model is updated with the obtained optical characterization data from the current sample and re-attempts the target until success. Table 2 displays the results for 5 single-wavelength targets. It took 2 attempts on average for a target to be successfully achieved. For each of the 10 samples grown during this stage, the model makes 5 predictions on its $\mathscr{R}$ for each wavelength. For these 50 total predictions, the mean absolute error (MAE) between the model predictions and measured results is 0.0246, which demonstrates the accuracy of the prediction. Moreover, the average model predicted uncertainty is 0.0267, and its proximity to the MAE demonstrates the accuracy of the model's estimate of its uncertainty. 

To further demonstrate the effectiveness of the implementation of calibration layers and autonomous learning, we perform a control experiment without these methods to benchmark the model prediction performance. In the control experiment, the model only goes through the predefined learning process, in which 17 samples are grown at $T$ between 820 and 880°C with 3.75°C increments, and for $t$ as specified in Eqn.(3). The system is then requested to produce silver thin films that satisfy the same 5 single-wavelength target. Despite the control experiment having the same amount of training sample data as the previous experiment, it requires on average 3.6 attempts to successfully achieve each target. The MAE between the model predictions and measured results is 0.0618, also significantly increased from the previous experiment (Figure 4(c)). The control experiment hence demonstrates the superior performance of the model using calibration layers and autonomous learning. 
\begin{table}[h]
    \centering
    \renewcommand{\arraystretch}{1.5} 
    \setlength{\tabcolsep}{2pt} 
    \resizebox{0.97\linewidth}{!}{ 
    \begin{tabular}{cccccccc}
    \toprule
    \textbf{Target} & \textbf{Attempt 1} & \textbf{2} & \textbf{3} & \textbf{4} & \textbf{5} & \textbf{6} & \textbf{7} \\
    \toprule
    $\mathscr{R}^{\text{target}}_{443} = 0.61$ & 0.609 &  &  &  &  &  &  \\
    \midrule
    $\mathscr{R}^{\text{target}}_{514} = 0.47$ & 0.413 & 0.452 &  &  &  &  &  \\
    \midrule
    $\mathscr{R}^{\text{target}}_{689} = 0.88$ & 0.891 &  &  &  &  &  &  \\
    \midrule
    $\mathscr{R}^{\text{target}}_{781} = 0.30$ & 0.298 &  &  &  &  &  &  \\
    \midrule
    $\mathscr{R}^{\text{target}}_{817} = 0.58$ & 0.545 & 0.636 & 0.541 & 0.510 & 0.585 &  &  \\
    \toprule
    $\mathscr{R}^{\text{target}}_{443} = 0.85$ & \multirow{2}{*}{\centering N/A} & \multirow{2}{*}{\centering N/A} & 0.914 & 0.845 &  &  &  \\
    $\mathscr{R}^{\text{target}}_{781} = 0.47$ &  &  & 0.654 & 0.507 &  &  &  \\
    \midrule
    $\mathscr{R}^{\text{target}}_{443} = 0.85$ & \multirow{2}{*}{\centering N/A} & \multirow{2}{*}{\centering N/A} & 0.797 & \multirow{2}{*}{\centering N/A} & 0.879 & 0.779 & 0.817 \\
    $\mathscr{R}^{\text{target}}_{781} = 0.35$ &  &  & 0.331 &  & 0.489 & 0.325 & 0.342 \\
    \toprule
    \end{tabular}
    }
    \captionsetup{justification=raggedright,singlelinecheck=false}
    \caption{\justifying{Results of achieving specified optical properties with single and multi-wavelength targets.}}
    \label{tab:example}
\end{table}

4.2 Multi-wavelength Target

We also specify $\mathscr{R}$ targets spanning multiple wavelengths. The loss function is defined as

\begin{equation}
\mathscr{L} = \sum_{\{\lambda\}_{\text{specified}}} (\mu_{\mathscr{R},\lambda} - \mathscr{R}_{\lambda}^{\text{target}})^2
\end{equation}

As the number of wavelengths specified in the target increases, it is not guaranteed that there exists a set of growth parameters $(T, t, \mathscr{R}_c)$ that could produce a film with the desired optical properties. Therefore, when the loss function is minimized and any of the $\mu_{\mathscr{R},\lambda}$'s is still $> 0.01$ away from the target, the algorithm decides that no growth parameters can be found to reach the target at the current sample's $\mathscr{R}_c$. The system aborts this sample and grows the calibration layer on the subsequent substrates until it finds one with a set of $(T, t, \mathscr{R}_c)$ that yields $\mu_{\mathscr{R},\lambda}$ within $\pm 0.01$ of the target for all specified wavelengths.  

The system proceeds to deposit the sample at the found $(T, t)$ and subsequently measures its optical properties $\{\mathscr{R}_{\lambda}, \mathscr{A}_{\lambda}\}$. The system then updates the model with all collected data. If  
\begin{equation}
\sum_{\{\lambda\}_{\text{specified}}} \left| \mathscr{R}_{\lambda} - \mathscr{R}_{\lambda}^{\text{target}} \right| < 0.025n
\end{equation}
where $n$ is the number of specified wavelengths in the target, the growth is considered success, and the system moves on to the next target. If unsuccessful, the system makes a new prediction using the updated model and re-attempts the target until success.

Table 2 displays the results using multi-wavelength targets. For certain samples, the model is unable to find a set of growth parameters that can yield a film to reach the specified multi-wavelength targets, denoted as N/A. These samples do not count towards the number of failed attempts. The system aborts the sample and proceeds to the next substrate. Upon switching to a sample with an $\mathscr{R}_c$ that leads to an achievable growth condition to obtain the desired optical properties, it begins the growth and can achieve the target within the allowed error. In this stage, 6 attempts are made to successfully achieve 2 targets. The success in achieving the two multi-wavelength targets demonstrates the versatility of our system to control the silver thin film's power-splitting spectrum over multiple wavelengths.

\section{Discussion and Conclusion}
In this work, we have established the ability of a fully autonomous ML-enabled system to address persisting challenges in the field of PVD thin film growth, namely the lengthy and labor-intensive human involvement and the difficulty of adapting to hidden parameters in the growth conditions. We see three components critical to a fully automated thin film deposition system: full mechanical and software automation, active learning for exploring the parameter space during data collection, and active targeting of desired outputs. Previous studies have implemented two of the three such as automation and active data collection \cite{noack2019kriging, noack2020autonomous} or manually-executed active training and target-seeking \cite{Wakabayashi2019}, but to our knowledge, none have performed all three. Our work demonstrates the first system to implement all three in the field of thin film deposition. Using Bayesian ML, we have engineered a fully autonomous system that learns the dynamics of silver growth while accounting for variations in substrates and chamber conditions, and reliably grows samples with desired optical properties.

We choose to work with silver thin films because they are well-understood but retain the difficult aspects of thin film growth. As such, the methods that we develop are broadly applicable. The procedure we demonstrate of using predefined learning to map the parameter space, autonomous learning to minimize model uncertainty, and optimization to seek specific targets is readily generalizable to other thin film growth tasks.

With the fully autonomous system, we grow a total of 38 samples in the training and testing stage and collect more than 20,000 data points without human intervention. This represents a significant advantage over human-led experimentation. This high-throughput operation is primarily enabled by our use of a 72-slot sample handing system and in-situ quasi-continuous optical characterization. By designing chambers for high-capacity sample transfer and measurement, similar throughput can be achieved for other thin film fabrication and in-situ characterization tasks.

In contrast to other ML techniques such as artificial neural networks, Bayesian optimization allows the quantification of prediction uncertainties. This aspect has been used to decide optimal points of parameter space exploration~\cite{Wakabayashi2019,Wakabayashi2022,Ohkubo2021}. Moreover, because of this ability to quantify uncertainty, our model sets numerical thresholds to determine both when to stop the learning process and when to conclude that a target is unattainable. By exploring the variety of decisions that the model is able to make, we further reduce the amount of human involvement in the process.

The high throughput of the system also enables the quantitative examination of the fluctuations in $\mathscr{R}$'s given the same growth parameters ($T$, $t$). This fluctuation is attributed to the hidden parameters in the growth process. We introduce the concept of the calibration layer that allows the model to partially capture the effect of these hidden parameters. With the information on the otherwise unknown hidden parameters, significant improvement in model prediction performance is achieved. The fluctuations in the growth process have been found in previous studies utilizing ML for thin film deposition~\cite{Febba2023,Ohkubo2021}, though its effect on the material property and model performance is often neglected. Shrivastava et al.\cite{shrivastava2024bayesian} have made an effort to deliberately avoid growing at growth parameters that are particularly sensitive to such fluctuations. Our work represents an important step toward quantifying these hidden parameters by incorporating the calibration layer measurement $\mathscr{R}_c$ into the model's parameter space.

One can extend the calibration layer approach by incorporating additional checkpoints along the deposition trajectories and optimize in a higher-dimensional parameter space. This could eventually lead to a quasi-continuous adaptive control algorithm where the system adjusts in real time based on feedback. However, implementing such adaptive control using reinforcement learning in complex experimental settings poses challenges, as it typically requires substantial amount of training trials \cite{Mnih2015Human}. Our method balances elements of adaptive control with efficient model training, accounting for both the limited data availability and the inherent variability in the material growth process. With more advanced reinforcement learning algorithms \cite{kamthe2018data} and higher sample throughput, one can envision that the quasi-continuous adaptive control may be implemented in thin film deposition in the future. 

\vspace{\baselineskip}
\noindent \textbf{Acknowledgments}

This work is supported by the National Science Foundation (NSF CNS-2019131) and University of Chicago Big Idea Generator Seed Grant. This work made use of the shared facilities at the University of Chicago Materials Research Science and Engineering Center, supported by National Science Foundation under award number DMR-2011854. This work made use of the Pritzker Nanofabrication Facility at the Pritzker School of Molecular Engineering at the University of Chicago, which receives support from Soft and Hybrid Nanotechnology Experimental (SHyNE) Resource (NSF ECCS-2025633), a node of the National Science Foundation’s National Nanotechnology Coordinated Infrastructure.

\setlength{\bibsep}{0pt}
\bibliography{references}

\begin{thebibliography}{36}%
\makeatletter
\providecommand \@ifxundefined [1]{%
 \@ifx{#1\undefined}
}%
\providecommand \@ifnum [1]{%
 \ifnum #1\expandafter \@firstoftwo
 \else \expandafter \@secondoftwo
 \fi
}%
\providecommand \@ifx [1]{%
 \ifx #1\expandafter \@firstoftwo
 \else \expandafter \@secondoftwo
 \fi
}%
\providecommand \natexlab [1]{#1}%
\providecommand \enquote  [1]{``#1''}%
\providecommand \bibnamefont  [1]{#1}%
\providecommand \bibfnamefont [1]{#1}%
\providecommand \citenamefont [1]{#1}%
\providecommand \href@noop [0]{\@secondoftwo}%
\providecommand \href [0]{\begingroup \@sanitize@url \@href}%
\providecommand \@href[1]{\@@startlink{#1}\@@href}%
\providecommand \@@href[1]{\endgroup#1\@@endlink}%
\providecommand \@sanitize@url [0]{\catcode `\\12\catcode `\$12\catcode `\&12\catcode `\#12\catcode `\^12\catcode `\_12\catcode `\%12\relax}%
\providecommand \@@startlink[1]{}%
\providecommand \@@endlink[0]{}%
\providecommand \url  [0]{\begingroup\@sanitize@url \@url }%
\providecommand \@url [1]{\endgroup\@href {#1}{\urlprefix }}%
\providecommand \urlprefix  [0]{URL }%
\providecommand \Eprint [0]{\href }%
\providecommand \doibase [0]{https://doi.org/}%
\providecommand \selectlanguage [0]{\@gobble}%
\providecommand \bibinfo  [0]{\@secondoftwo}%
\providecommand \bibfield  [0]{\@secondoftwo}%
\providecommand \translation [1]{[#1]}%
\providecommand \BibitemOpen [0]{}%
\providecommand \bibitemStop [0]{}%
\providecommand \bibitemNoStop [0]{.\EOS\space}%
\providecommand \EOS [0]{\spacefactor3000\relax}%
\providecommand \BibitemShut  [1]{\csname bibitem#1\endcsname}%
\let\auto@bib@innerbib\@empty
\bibitem [{\citenamefont {Jordan}\ and\ \citenamefont {Mitchell}(2015)}]{Jordan2015}%
  \BibitemOpen
  \bibfield  {author} {\bibinfo {author} {\bibfnamefont {M.~I.}\ \bibnamefont {Jordan}}\ and\ \bibinfo {author} {\bibfnamefont {T.~M.}\ \bibnamefont {Mitchell}},\ }\bibfield  {title} {\bibinfo {title} {Machine learning: Trends, perspectives, and prospects},\ }\href@noop {} {\bibfield  {journal} {\bibinfo  {journal} {Science}\ }\textbf {\bibinfo {volume} {349}},\ \bibinfo {pages} {255} (\bibinfo {year} {2015})}\BibitemShut {NoStop}%
\bibitem [{\citenamefont {Garcia}\ \emph {et~al.}(2007)\citenamefont {Garcia}, \citenamefont {Jimenez}, \citenamefont {De~Santos},\ and\ \citenamefont {Armada}}]{Garcia2007}%
  \BibitemOpen
  \bibfield  {author} {\bibinfo {author} {\bibfnamefont {E.}~\bibnamefont {Garcia}}, \bibinfo {author} {\bibfnamefont {M.~A.}\ \bibnamefont {Jimenez}}, \bibinfo {author} {\bibfnamefont {P.~G.}\ \bibnamefont {De~Santos}},\ and\ \bibinfo {author} {\bibfnamefont {M.}~\bibnamefont {Armada}},\ }\bibfield  {title} {\bibinfo {title} {The evolution of robotics research},\ }\href@noop {} {\bibfield  {journal} {\bibinfo  {journal} {IEEE Robotics I\& Automation Magazine}\ }\textbf {\bibinfo {volume} {14}},\ \bibinfo {pages} {90} (\bibinfo {year} {2007})}\BibitemShut {NoStop}%
\bibitem [{\citenamefont {Morgan}\ and\ \citenamefont {Jacobs}(2020)}]{Morgan2020}%
  \BibitemOpen
  \bibfield  {author} {\bibinfo {author} {\bibfnamefont {D.}~\bibnamefont {Morgan}}\ and\ \bibinfo {author} {\bibfnamefont {R.}~\bibnamefont {Jacobs}},\ }\bibfield  {title} {\bibinfo {title} {Opportunities and challenges for machine learning in materials science},\ }\href@noop {} {\bibfield  {journal} {\bibinfo  {journal} {Annual Review of Materials Research}\ }\textbf {\bibinfo {volume} {50}},\ \bibinfo {pages} {71} (\bibinfo {year} {2020})}\BibitemShut {NoStop}%
\bibitem [{\citenamefont {Wakabayashi}\ \emph {et~al.}(2023)\citenamefont {Wakabayashi}, \citenamefont {Otsuka}, \citenamefont {Krockenberger}, \citenamefont {Sawada}, \citenamefont {Taniyasu},\ and\ \citenamefont {Yamamoto}}]{Wakabayashi2023}%
  \BibitemOpen
  \bibfield  {author} {\bibinfo {author} {\bibfnamefont {Y.~K.}\ \bibnamefont {Wakabayashi}}, \bibinfo {author} {\bibfnamefont {T.}~\bibnamefont {Otsuka}}, \bibinfo {author} {\bibfnamefont {Y.}~\bibnamefont {Krockenberger}}, \bibinfo {author} {\bibfnamefont {H.}~\bibnamefont {Sawada}}, \bibinfo {author} {\bibfnamefont {Y.}~\bibnamefont {Taniyasu}},\ and\ \bibinfo {author} {\bibfnamefont {H.}~\bibnamefont {Yamamoto}},\ }\bibfield  {title} {\bibinfo {title} {Stoichiometric growth of srtio3 films via bayesian optimization with adaptive prior mean},\ }\href@noop {} {\bibfield  {journal} {\bibinfo  {journal} {APL Machine Learning}\ }\textbf {\bibinfo {volume} {1}} (\bibinfo {year} {2023})}\BibitemShut {NoStop}%
\bibitem [{\citenamefont {Fébba}\ \emph {et~al.}(2023)\citenamefont {Fébba}, \citenamefont {Talley}, \citenamefont {Johnson}, \citenamefont {Schaefer}, \citenamefont {Bauers}, \citenamefont {Mangum},\ and\ \citenamefont {Zakutayev}}]{Febba2023}%
  \BibitemOpen
  \bibfield  {author} {\bibinfo {author} {\bibfnamefont {D.~M.}\ \bibnamefont {Fébba}}, \bibinfo {author} {\bibfnamefont {K.~R.}\ \bibnamefont {Talley}}, \bibinfo {author} {\bibfnamefont {K.}~\bibnamefont {Johnson}}, \bibinfo {author} {\bibfnamefont {S.}~\bibnamefont {Schaefer}}, \bibinfo {author} {\bibfnamefont {S.~R.}\ \bibnamefont {Bauers}}, \bibinfo {author} {\bibfnamefont {J.~S.}\ \bibnamefont {Mangum}},\ and\ \bibinfo {author} {\bibfnamefont {A.}~\bibnamefont {Zakutayev}},\ }\bibfield  {title} {\bibinfo {title} {Autonomous sputter synthesis of thin film nitrides with composition controlled by bayesian optimization of optical plasma emission},\ }\href@noop {} {\bibfield  {journal} {\bibinfo  {journal} {APL Materials}\ }\textbf {\bibinfo {volume} {11}},\ \bibinfo {pages} {7} (\bibinfo {year} {2023})}\BibitemShut {NoStop}%
\bibitem [{\citenamefont {Johnson}\ \emph {et~al.}(2024)\citenamefont {Johnson}, \citenamefont {Mishra}, \citenamefont {Kirsch},\ and\ \citenamefont {Mehta}}]{Johnson2024Active}%
  \BibitemOpen
  \bibfield  {author} {\bibinfo {author} {\bibfnamefont {N.~S.}\ \bibnamefont {Johnson}}, \bibinfo {author} {\bibfnamefont {A.~A.}\ \bibnamefont {Mishra}}, \bibinfo {author} {\bibfnamefont {D.~J.}\ \bibnamefont {Kirsch}},\ and\ \bibinfo {author} {\bibfnamefont {A.}~\bibnamefont {Mehta}},\ }\bibfield  {title} {\bibinfo {title} {Active learning for rapid targeted synthesis of compositionally complex alloys},\ }\href@noop {} {\bibfield  {journal} {\bibinfo  {journal} {Materials}\ }\textbf {\bibinfo {volume} {17}},\ \bibinfo {pages} {4038} (\bibinfo {year} {2024})}\BibitemShut {NoStop}%
\bibitem [{\citenamefont {Ishiyama}\ \emph {et~al.}(2024)\citenamefont {Ishiyama}, \citenamefont {Nozawa}, \citenamefont {Nishida}, \citenamefont {Suemasu},\ and\ \citenamefont {Toko}}]{ishiyama2024bayesian}%
  \BibitemOpen
  \bibfield  {author} {\bibinfo {author} {\bibfnamefont {T.}~\bibnamefont {Ishiyama}}, \bibinfo {author} {\bibfnamefont {K.}~\bibnamefont {Nozawa}}, \bibinfo {author} {\bibfnamefont {T.}~\bibnamefont {Nishida}}, \bibinfo {author} {\bibfnamefont {T.}~\bibnamefont {Suemasu}},\ and\ \bibinfo {author} {\bibfnamefont {K.}~\bibnamefont {Toko}},\ }\bibfield  {title} {\bibinfo {title} {Bayesian optimization-driven enhancement of the thermoelectric properties of polycrystalline iii-v semiconductor thin films},\ }\href@noop {} {\bibfield  {journal} {\bibinfo  {journal} {NPG Asia Materials}\ }\textbf {\bibinfo {volume} {16}},\ \bibinfo {pages} {17} (\bibinfo {year} {2024})}\BibitemShut {NoStop}%
\bibitem [{\citenamefont {Shrivastava}\ \emph {et~al.}(2024)\citenamefont {Shrivastava}, \citenamefont {Kalaswad}, \citenamefont {Custer}, \citenamefont {Adams},\ and\ \citenamefont {Najm}}]{shrivastava2024bayesian}%
  \BibitemOpen
  \bibfield  {author} {\bibinfo {author} {\bibfnamefont {A.}~\bibnamefont {Shrivastava}}, \bibinfo {author} {\bibfnamefont {M.}~\bibnamefont {Kalaswad}}, \bibinfo {author} {\bibfnamefont {J.~O.}\ \bibnamefont {Custer}}, \bibinfo {author} {\bibfnamefont {D.~P.}\ \bibnamefont {Adams}},\ and\ \bibinfo {author} {\bibfnamefont {H.~N.}\ \bibnamefont {Najm}},\ }\bibfield  {title} {\bibinfo {title} {Bayesian optimization for stable properties amid processing fluctuations in sputter deposition},\ }\href@noop {} {\bibfield  {journal} {\bibinfo  {journal} {Journal of Vacuum Science \& Technology A}\ }\textbf {\bibinfo {volume} {42}} (\bibinfo {year} {2024})}\BibitemShut {NoStop}%
\bibitem [{\citenamefont {Wakabayashi}\ \emph {et~al.}(2019)\citenamefont {Wakabayashi}, \citenamefont {Otsuka}, \citenamefont {Krockenberger}, \citenamefont {Sawada}, \citenamefont {Taniyasu},\ and\ \citenamefont {Yamamoto}}]{Wakabayashi2019}%
  \BibitemOpen
  \bibfield  {author} {\bibinfo {author} {\bibfnamefont {Y.~K.}\ \bibnamefont {Wakabayashi}}, \bibinfo {author} {\bibfnamefont {T.}~\bibnamefont {Otsuka}}, \bibinfo {author} {\bibfnamefont {Y.}~\bibnamefont {Krockenberger}}, \bibinfo {author} {\bibfnamefont {H.}~\bibnamefont {Sawada}}, \bibinfo {author} {\bibfnamefont {Y.}~\bibnamefont {Taniyasu}},\ and\ \bibinfo {author} {\bibfnamefont {H.}~\bibnamefont {Yamamoto}},\ }\bibfield  {title} {\bibinfo {title} {Machine-learning-assisted thin-film growth: Bayesian optimization in molecular beam epitaxy of srruo3 thin films},\ }\href@noop {} {\bibfield  {journal} {\bibinfo  {journal} {APL Materials}\ }\textbf {\bibinfo {volume} {7}},\ \bibinfo {pages} {10} (\bibinfo {year} {2019})}\BibitemShut {NoStop}%
\bibitem [{\citenamefont {Messecar}\ \emph {et~al.}(2024)\citenamefont {Messecar}, \citenamefont {Durbin},\ and\ \citenamefont {Makin}}]{messecar2024quantum}%
  \BibitemOpen
  \bibfield  {author} {\bibinfo {author} {\bibfnamefont {A.~S.}\ \bibnamefont {Messecar}}, \bibinfo {author} {\bibfnamefont {S.~M.}\ \bibnamefont {Durbin}},\ and\ \bibinfo {author} {\bibfnamefont {R.~A.}\ \bibnamefont {Makin}},\ }\bibfield  {title} {\bibinfo {title} {Quantum and classical machine learning investigation of synthesis–structure relationships in epitaxially grown wide band gap semiconductors},\ }\href@noop {} {\bibfield  {journal} {\bibinfo  {journal} {MRS Communications}\ }\textbf {\bibinfo {volume} {14}},\ \bibinfo {pages} {660} (\bibinfo {year} {2024})}\BibitemShut {NoStop}%
\bibitem [{\citenamefont {Shen}\ \emph {et~al.}(2024)\citenamefont {Shen}, \citenamefont {Zhan}, \citenamefont {Xin}, \citenamefont {Li}, \citenamefont {Sun}, \citenamefont {Cong}, \citenamefont {Xu}, \citenamefont {Tang}, \citenamefont {Wu}, \citenamefont {Xu},\ and\ \citenamefont {Wei}}]{shen2024machine}%
  \BibitemOpen
  \bibfield  {author} {\bibinfo {author} {\bibfnamefont {C.}~\bibnamefont {Shen}}, \bibinfo {author} {\bibfnamefont {W.}~\bibnamefont {Zhan}}, \bibinfo {author} {\bibfnamefont {K.}~\bibnamefont {Xin}}, \bibinfo {author} {\bibfnamefont {M.}~\bibnamefont {Li}}, \bibinfo {author} {\bibfnamefont {Z.}~\bibnamefont {Sun}}, \bibinfo {author} {\bibfnamefont {H.}~\bibnamefont {Cong}}, \bibinfo {author} {\bibfnamefont {C.}~\bibnamefont {Xu}}, \bibinfo {author} {\bibfnamefont {J.}~\bibnamefont {Tang}}, \bibinfo {author} {\bibfnamefont {Z.}~\bibnamefont {Wu}}, \bibinfo {author} {\bibfnamefont {B.}~\bibnamefont {Xu}},\ and\ \bibinfo {author} {\bibfnamefont {Z.}~\bibnamefont {Wei}},\ }\bibfield  {title} {\bibinfo {title} {Machine-learning-assisted and real-time-feedback-controlled growth of inas/gaas quantum dots},\ }\href@noop {} {\bibfield  {journal} {\bibinfo  {journal} {Nature Communications}\ }\textbf {\bibinfo {volume} {15}},\ \bibinfo {pages} {2724} (\bibinfo {year} {2024})}\BibitemShut {NoStop}%
\bibitem [{\citenamefont {Kim}\ \emph {et~al.}(2023)\citenamefont {Kim}, \citenamefont {Chong}, \citenamefont {Rhee}, \citenamefont {Khim}, \citenamefont {Jung}, \citenamefont {Kim}, \citenamefont {Jeong}, \citenamefont {Choi},\ and\ \citenamefont {Chang}}]{kim2023machine}%
  \BibitemOpen
  \bibfield  {author} {\bibinfo {author} {\bibfnamefont {H.~J.}\ \bibnamefont {Kim}}, \bibinfo {author} {\bibfnamefont {M.}~\bibnamefont {Chong}}, \bibinfo {author} {\bibfnamefont {T.~G.}\ \bibnamefont {Rhee}}, \bibinfo {author} {\bibfnamefont {Y.~G.}\ \bibnamefont {Khim}}, \bibinfo {author} {\bibfnamefont {M.~H.}\ \bibnamefont {Jung}}, \bibinfo {author} {\bibfnamefont {Y.~M.}\ \bibnamefont {Kim}}, \bibinfo {author} {\bibfnamefont {H.~Y.}\ \bibnamefont {Jeong}}, \bibinfo {author} {\bibfnamefont {B.~K.}\ \bibnamefont {Choi}},\ and\ \bibinfo {author} {\bibfnamefont {Y.~J.}\ \bibnamefont {Chang}},\ }\bibfield  {title} {\bibinfo {title} {Machine-learning-assisted analysis of transition metal dichalcogenide thin-film growth},\ }\href@noop {} {\bibfield  {journal} {\bibinfo  {journal} {Nano Convergence}\ }\textbf {\bibinfo {volume} {10}},\ \bibinfo {pages} {10} (\bibinfo {year} {2023})}\BibitemShut {NoStop}%
\bibitem [{\citenamefont {Provence}\ \emph {et~al.}(2020)\citenamefont {Provence}, \citenamefont {Thapa}, \citenamefont {Paudel}, \citenamefont {Truttmann}, \citenamefont {Prakash}, \citenamefont {Jalan},\ and\ \citenamefont {Comes}}]{provence2020machine}%
  \BibitemOpen
  \bibfield  {author} {\bibinfo {author} {\bibfnamefont {S.~R.}\ \bibnamefont {Provence}}, \bibinfo {author} {\bibfnamefont {S.}~\bibnamefont {Thapa}}, \bibinfo {author} {\bibfnamefont {R.}~\bibnamefont {Paudel}}, \bibinfo {author} {\bibfnamefont {T.~K.}\ \bibnamefont {Truttmann}}, \bibinfo {author} {\bibfnamefont {A.}~\bibnamefont {Prakash}}, \bibinfo {author} {\bibfnamefont {B.}~\bibnamefont {Jalan}},\ and\ \bibinfo {author} {\bibfnamefont {R.~B.}\ \bibnamefont {Comes}},\ }\bibfield  {title} {\bibinfo {title} {Machine learning analysis of perovskite oxides grown by molecular beam epitaxy},\ }\href@noop {} {\bibfield  {journal} {\bibinfo  {journal} {Physical Review Materials}\ }\textbf {\bibinfo {volume} {4}},\ \bibinfo {pages} {083807} (\bibinfo {year} {2020})}\BibitemShut {NoStop}%
\bibitem [{\citenamefont {Guevarra}\ \emph {et~al.}(2022)\citenamefont {Guevarra}, \citenamefont {Zhou}, \citenamefont {Richter}, \citenamefont {Shinde}, \citenamefont {Chen}, \citenamefont {Gomes},\ and\ \citenamefont {Gregoire}}]{Guevarra2022Materials}%
  \BibitemOpen
  \bibfield  {author} {\bibinfo {author} {\bibfnamefont {D.}~\bibnamefont {Guevarra}}, \bibinfo {author} {\bibfnamefont {L.}~\bibnamefont {Zhou}}, \bibinfo {author} {\bibfnamefont {M.~H.}\ \bibnamefont {Richter}}, \bibinfo {author} {\bibfnamefont {A.}~\bibnamefont {Shinde}}, \bibinfo {author} {\bibfnamefont {D.}~\bibnamefont {Chen}}, \bibinfo {author} {\bibfnamefont {C.~P.}\ \bibnamefont {Gomes}},\ and\ \bibinfo {author} {\bibfnamefont {J.~M.}\ \bibnamefont {Gregoire}},\ }\bibfield  {title} {\bibinfo {title} {Materials structure–property factorization for identification of synergistic phase interactions in complex solar fuels photoanodes},\ }\href@noop {} {\bibfield  {journal} {\bibinfo  {journal} {npj Computational Materials}\ }\textbf {\bibinfo {volume} {8}},\ \bibinfo {pages} {57} (\bibinfo {year} {2022})}\BibitemShut {NoStop}%
\bibitem [{\citenamefont {Ni}\ and\ \citenamefont {Matsui}(2022)}]{Ni2022Phase}%
  \BibitemOpen
  \bibfield  {author} {\bibinfo {author} {\bibfnamefont {Z.}~\bibnamefont {Ni}}\ and\ \bibinfo {author} {\bibfnamefont {H.}~\bibnamefont {Matsui}},\ }\bibfield  {title} {\bibinfo {title} {Phase control of heterogeneous {Hf}\textsubscript{X}{Zr}\textsubscript{(1-X)}{O}\textsubscript{2} thin films by machine learning},\ }\href@noop {} {\bibfield  {journal} {\bibinfo  {journal} {Japanese Journal of Applied Physics}\ }\textbf {\bibinfo {volume} {61}},\ \bibinfo {pages} {SH1009} (\bibinfo {year} {2022})}\BibitemShut {NoStop}%
\bibitem [{\citenamefont {Liang}\ \emph {et~al.}(2022)\citenamefont {Liang}, \citenamefont {Stanev}, \citenamefont {Kusne}, \citenamefont {Tsukahara}, \citenamefont {Ito}, \citenamefont {Takahashi}, \citenamefont {Lippmaa},\ and\ \citenamefont {Takeuchi}}]{Liang2022Application}%
  \BibitemOpen
  \bibfield  {author} {\bibinfo {author} {\bibfnamefont {H.}~\bibnamefont {Liang}}, \bibinfo {author} {\bibfnamefont {V.}~\bibnamefont {Stanev}}, \bibinfo {author} {\bibfnamefont {A.~G.}\ \bibnamefont {Kusne}}, \bibinfo {author} {\bibfnamefont {Y.}~\bibnamefont {Tsukahara}}, \bibinfo {author} {\bibfnamefont {K.}~\bibnamefont {Ito}}, \bibinfo {author} {\bibfnamefont {R.}~\bibnamefont {Takahashi}}, \bibinfo {author} {\bibfnamefont {M.}~\bibnamefont {Lippmaa}},\ and\ \bibinfo {author} {\bibfnamefont {I.}~\bibnamefont {Takeuchi}},\ }\bibfield  {title} {\bibinfo {title} {Application of machine learning to reflection high-energy electron diffraction images for automated structural phase mapping},\ }\href@noop {} {\bibfield  {journal} {\bibinfo  {journal} {Physical Review Materials}\ }\textbf {\bibinfo {volume} {6}},\ \bibinfo {pages} {063805} (\bibinfo {year} {2022})}\BibitemShut {NoStop}%
\bibitem [{\citenamefont {Ament}\ \emph {et~al.}(2021)\citenamefont {Ament}, \citenamefont {Amsler}, \citenamefont {Sutherland}, \citenamefont {Chang}, \citenamefont {Guevarra}, \citenamefont {Connolly}, \citenamefont {Gregoire}, \citenamefont {Thompson}, \citenamefont {Gomes},\ and\ \citenamefont {Van~Dover}}]{Ament2021Autonomous}%
  \BibitemOpen
  \bibfield  {author} {\bibinfo {author} {\bibfnamefont {S.}~\bibnamefont {Ament}}, \bibinfo {author} {\bibfnamefont {M.}~\bibnamefont {Amsler}}, \bibinfo {author} {\bibfnamefont {D.~R.}\ \bibnamefont {Sutherland}}, \bibinfo {author} {\bibfnamefont {M.~C.}\ \bibnamefont {Chang}}, \bibinfo {author} {\bibfnamefont {D.}~\bibnamefont {Guevarra}}, \bibinfo {author} {\bibfnamefont {A.~B.}\ \bibnamefont {Connolly}}, \bibinfo {author} {\bibfnamefont {J.~M.}\ \bibnamefont {Gregoire}}, \bibinfo {author} {\bibfnamefont {M.~O.}\ \bibnamefont {Thompson}}, \bibinfo {author} {\bibfnamefont {C.~P.}\ \bibnamefont {Gomes}},\ and\ \bibinfo {author} {\bibfnamefont {R.~B.}\ \bibnamefont {Van~Dover}},\ }\bibfield  {title} {\bibinfo {title} {Autonomous materials synthesis via hierarchical active learning of nonequilibrium phase diagrams},\ }\href@noop {} {\bibfield  {journal} {\bibinfo  {journal} {Science Advances}\ }\textbf {\bibinfo {volume} {7}},\ \bibinfo {pages} {eabg4930} (\bibinfo {year} {2021})}\BibitemShut {NoStop}%
\bibitem [{\citenamefont {Ohkubo}\ \emph {et~al.}(2021)\citenamefont {Ohkubo}, \citenamefont {Hou}, \citenamefont {Lee}, \citenamefont {Aizawa}, \citenamefont {Lippmaa}, \citenamefont {Chikyow},\ and\ \citenamefont {Mori}}]{Ohkubo2021}%
  \BibitemOpen
  \bibfield  {author} {\bibinfo {author} {\bibfnamefont {I.}~\bibnamefont {Ohkubo}}, \bibinfo {author} {\bibfnamefont {Z.}~\bibnamefont {Hou}}, \bibinfo {author} {\bibfnamefont {J.~N.}\ \bibnamefont {Lee}}, \bibinfo {author} {\bibfnamefont {T.}~\bibnamefont {Aizawa}}, \bibinfo {author} {\bibfnamefont {M.}~\bibnamefont {Lippmaa}}, \bibinfo {author} {\bibfnamefont {T.}~\bibnamefont {Chikyow}},\ and\ \bibinfo {author} {\bibfnamefont {T.}~\bibnamefont {Mori}},\ }\bibfield  {title} {\bibinfo {title} {Realization of closed-loop optimization of epitaxial titanium nitride thin-film growth via machine learning},\ }\href@noop {} {\bibfield  {journal} {\bibinfo  {journal} {Materials Today Physics}\ }\textbf {\bibinfo {volume} {16}},\ \bibinfo {pages} {100296} (\bibinfo {year} {2021})}\BibitemShut {NoStop}%
\bibitem [{\citenamefont {Wakabayashi}\ \emph {et~al.}(2022)\citenamefont {Wakabayashi}, \citenamefont {Otsuka}, \citenamefont {Krockenberger}, \citenamefont {Sawada}, \citenamefont {Taniyasu},\ and\ \citenamefont {Yamamoto}}]{Wakabayashi2022}%
  \BibitemOpen
  \bibfield  {author} {\bibinfo {author} {\bibfnamefont {Y.~K.}\ \bibnamefont {Wakabayashi}}, \bibinfo {author} {\bibfnamefont {T.}~\bibnamefont {Otsuka}}, \bibinfo {author} {\bibfnamefont {Y.}~\bibnamefont {Krockenberger}}, \bibinfo {author} {\bibfnamefont {H.}~\bibnamefont {Sawada}}, \bibinfo {author} {\bibfnamefont {Y.}~\bibnamefont {Taniyasu}},\ and\ \bibinfo {author} {\bibfnamefont {H.}~\bibnamefont {Yamamoto}},\ }\bibfield  {title} {\bibinfo {title} {Bayesian optimization with experimental failure for high-throughput materials growth},\ }\href@noop {} {\bibfield  {journal} {\bibinfo  {journal} {npj Computational Materials}\ }\textbf {\bibinfo {volume} {8}},\ \bibinfo {pages} {180} (\bibinfo {year} {2022})}\BibitemShut {NoStop}%
\bibitem [{\citenamefont {Packwood}(2017)}]{Packwood2017Bayesian}%
  \BibitemOpen
  \bibfield  {author} {\bibinfo {author} {\bibfnamefont {D.}~\bibnamefont {Packwood}},\ }\href@noop {} {\emph {\bibinfo {title} {Bayesian Optimization for Materials Science}}}\ (\bibinfo  {publisher} {Springer Singapore},\ \bibinfo {address} {Singapore},\ \bibinfo {year} {2017})\BibitemShut {NoStop}%
\bibitem [{\citenamefont {Shahriari}\ \emph {et~al.}(2015)\citenamefont {Shahriari}, \citenamefont {Swersky}, \citenamefont {Wang}, \citenamefont {Adams},\ and\ \citenamefont {De~Freitas}}]{Shahriari2015}%
  \BibitemOpen
  \bibfield  {author} {\bibinfo {author} {\bibfnamefont {B.}~\bibnamefont {Shahriari}}, \bibinfo {author} {\bibfnamefont {K.}~\bibnamefont {Swersky}}, \bibinfo {author} {\bibfnamefont {Z.}~\bibnamefont {Wang}}, \bibinfo {author} {\bibfnamefont {R.~P.}\ \bibnamefont {Adams}},\ and\ \bibinfo {author} {\bibfnamefont {N.}~\bibnamefont {De~Freitas}},\ }\bibfield  {title} {\bibinfo {title} {Taking the human out of the loop: A review of bayesian optimization},\ }\href@noop {} {\bibfield  {journal} {\bibinfo  {journal} {Proceedings of the IEEE}\ }\textbf {\bibinfo {volume} {104}},\ \bibinfo {pages} {148} (\bibinfo {year} {2015})}\BibitemShut {NoStop}%
\bibitem [{\citenamefont {Kusne}\ \emph {et~al.}(2020)\citenamefont {Kusne}, \citenamefont {Yu}, \citenamefont {Wu}, \citenamefont {Zhang}, \citenamefont {Hattrick-Simpers}, \citenamefont {DeCost}, \citenamefont {Sarker}, \citenamefont {Oses}, \citenamefont {Toher}, \citenamefont {Curtarolo},\ and\ \citenamefont {Davydov}}]{kusne2020on}%
  \BibitemOpen
  \bibfield  {author} {\bibinfo {author} {\bibfnamefont {A.~G.}\ \bibnamefont {Kusne}}, \bibinfo {author} {\bibfnamefont {H.}~\bibnamefont {Yu}}, \bibinfo {author} {\bibfnamefont {C.}~\bibnamefont {Wu}}, \bibinfo {author} {\bibfnamefont {H.}~\bibnamefont {Zhang}}, \bibinfo {author} {\bibfnamefont {J.}~\bibnamefont {Hattrick-Simpers}}, \bibinfo {author} {\bibfnamefont {B.}~\bibnamefont {DeCost}}, \bibinfo {author} {\bibfnamefont {S.}~\bibnamefont {Sarker}}, \bibinfo {author} {\bibfnamefont {C.}~\bibnamefont {Oses}}, \bibinfo {author} {\bibfnamefont {C.}~\bibnamefont {Toher}}, \bibinfo {author} {\bibfnamefont {S.}~\bibnamefont {Curtarolo}},\ and\ \bibinfo {author} {\bibfnamefont {A.~V.}\ \bibnamefont {Davydov}},\ }\bibfield  {title} {\bibinfo {title} {On-the-fly closed-loop materials discovery via bayesian active learning},\ }\href@noop {} {\bibfield  {journal} {\bibinfo  {journal} {Nature Communications}\ }\textbf {\bibinfo {volume} {11}},\ \bibinfo {pages} {5966} (\bibinfo {year} {2020})}\BibitemShut
  {NoStop}%
\bibitem [{\citenamefont {Shimizu}\ \emph {et~al.}(2020)\citenamefont {Shimizu}, \citenamefont {Kobayashi}, \citenamefont {Watanabe}, \citenamefont {Ando},\ and\ \citenamefont {Hitosugi}}]{Shimizu2020}%
  \BibitemOpen
  \bibfield  {author} {\bibinfo {author} {\bibfnamefont {R.}~\bibnamefont {Shimizu}}, \bibinfo {author} {\bibfnamefont {S.}~\bibnamefont {Kobayashi}}, \bibinfo {author} {\bibfnamefont {Y.}~\bibnamefont {Watanabe}}, \bibinfo {author} {\bibfnamefont {Y.}~\bibnamefont {Ando}},\ and\ \bibinfo {author} {\bibfnamefont {T.}~\bibnamefont {Hitosugi}},\ }\bibfield  {title} {\bibinfo {title} {Autonomous materials synthesis by machine learning and robotics},\ }\href@noop {} {\bibfield  {journal} {\bibinfo  {journal} {APL Materials}\ }\textbf {\bibinfo {volume} {8}},\ \bibinfo {pages} {11} (\bibinfo {year} {2020})}\BibitemShut {NoStop}%
\bibitem [{\citenamefont {Zhao}\ \emph {et~al.}(2009)\citenamefont {Zhao}, \citenamefont {Su}, \citenamefont {Wang}, \citenamefont {Xu},\ and\ \citenamefont {Zhang}}]{Zhao2009}%
  \BibitemOpen
  \bibfield  {author} {\bibinfo {author} {\bibfnamefont {P.}~\bibnamefont {Zhao}}, \bibinfo {author} {\bibfnamefont {W.}~\bibnamefont {Su}}, \bibinfo {author} {\bibfnamefont {R.}~\bibnamefont {Wang}}, \bibinfo {author} {\bibfnamefont {X.}~\bibnamefont {Xu}},\ and\ \bibinfo {author} {\bibfnamefont {F.}~\bibnamefont {Zhang}},\ }\bibfield  {title} {\bibinfo {title} {Properties of thin silver films with different thickness},\ }\href@noop {} {\bibfield  {journal} {\bibinfo  {journal} {Physica E: Low-dimensional Systems and Nanostructures}\ }\textbf {\bibinfo {volume} {41}},\ \bibinfo {pages} {387} (\bibinfo {year} {2009})}\BibitemShut {NoStop}%
\bibitem [{\citenamefont {Savaloni}\ and\ \citenamefont {Firouzi-Arani}(2008)}]{Savaloni2008}%
  \BibitemOpen
  \bibfield  {author} {\bibinfo {author} {\bibfnamefont {H.}~\bibnamefont {Savaloni}}\ and\ \bibinfo {author} {\bibfnamefont {M.}~\bibnamefont {Firouzi-Arani}},\ }\bibfield  {title} {\bibinfo {title} {Dependence of the optical properties of uhv deposited silver thin films on the deposition parameters and their relation to the nanostructure of the films},\ }\href@noop {} {\bibfield  {journal} {\bibinfo  {journal} {Philosophical Magazine}\ }\textbf {\bibinfo {volume} {88}},\ \bibinfo {pages} {711} (\bibinfo {year} {2008})}\BibitemShut {NoStop}%
\bibitem [{\citenamefont {Reddy}\ \emph {et~al.}(2017)\citenamefont {Reddy}, \citenamefont {Guler}, \citenamefont {Chaudhuri}, \citenamefont {Dutta}, \citenamefont {Kildishev}, \citenamefont {Shalaev},\ and\ \citenamefont {Boltasseva}}]{Reddy2017}%
  \BibitemOpen
  \bibfield  {author} {\bibinfo {author} {\bibfnamefont {H.}~\bibnamefont {Reddy}}, \bibinfo {author} {\bibfnamefont {U.}~\bibnamefont {Guler}}, \bibinfo {author} {\bibfnamefont {K.}~\bibnamefont {Chaudhuri}}, \bibinfo {author} {\bibfnamefont {A.}~\bibnamefont {Dutta}}, \bibinfo {author} {\bibfnamefont {A.~V.}\ \bibnamefont {Kildishev}}, \bibinfo {author} {\bibfnamefont {V.~M.}\ \bibnamefont {Shalaev}},\ and\ \bibinfo {author} {\bibfnamefont {A.}~\bibnamefont {Boltasseva}},\ }\bibfield  {title} {\bibinfo {title} {Temperature-dependent optical properties of single crystalline and polycrystalline silver thin films},\ }\href@noop {} {\bibfield  {journal} {\bibinfo  {journal} {ACS Photonics}\ }\textbf {\bibinfo {volume} {4}},\ \bibinfo {pages} {1083} (\bibinfo {year} {2017})}\BibitemShut {NoStop}%
\bibitem [{\citenamefont {Choi}\ \emph {et~al.}(2020)\citenamefont {Choi}, \citenamefont {Cheng}, \citenamefont {Cleary}, \citenamefont {Sun}, \citenamefont {Dass}, \citenamefont {Hendrickson},\ and\ \citenamefont {Li}}]{Choi2020}%
  \BibitemOpen
  \bibfield  {author} {\bibinfo {author} {\bibfnamefont {J.}~\bibnamefont {Choi}}, \bibinfo {author} {\bibfnamefont {F.}~\bibnamefont {Cheng}}, \bibinfo {author} {\bibfnamefont {J.~W.}\ \bibnamefont {Cleary}}, \bibinfo {author} {\bibfnamefont {L.}~\bibnamefont {Sun}}, \bibinfo {author} {\bibfnamefont {C.~K.}\ \bibnamefont {Dass}}, \bibinfo {author} {\bibfnamefont {J.~R.}\ \bibnamefont {Hendrickson}},\ and\ \bibinfo {author} {\bibfnamefont {X.}~\bibnamefont {Li}},\ }\bibfield  {title} {\bibinfo {title} {Optical dielectric constants of single crystalline silver films in the long wavelength range},\ }\href@noop {} {\bibfield  {journal} {\bibinfo  {journal} {Optical Materials Express}\ }\textbf {\bibinfo {volume} {10}},\ \bibinfo {pages} {693} (\bibinfo {year} {2020})}\BibitemShut {NoStop}%
\bibitem [{\citenamefont {Faeth}\ \emph {et~al.}(2021)\citenamefont {Faeth}, \citenamefont {Yang}, \citenamefont {Kawasaki}, \citenamefont {Nelson}, \citenamefont {Mishra}, \citenamefont {Parzyck}, \citenamefont {Li}, \citenamefont {Schlom},\ and\ \citenamefont {Shen}}]{PhysRevX.11.021054}%
  \BibitemOpen
  \bibfield  {author} {\bibinfo {author} {\bibfnamefont {B.~D.}\ \bibnamefont {Faeth}}, \bibinfo {author} {\bibfnamefont {S.-L.}\ \bibnamefont {Yang}}, \bibinfo {author} {\bibfnamefont {J.~K.}\ \bibnamefont {Kawasaki}}, \bibinfo {author} {\bibfnamefont {J.~N.}\ \bibnamefont {Nelson}}, \bibinfo {author} {\bibfnamefont {P.}~\bibnamefont {Mishra}}, \bibinfo {author} {\bibfnamefont {C.~T.}\ \bibnamefont {Parzyck}}, \bibinfo {author} {\bibfnamefont {C.}~\bibnamefont {Li}}, \bibinfo {author} {\bibfnamefont {D.~G.}\ \bibnamefont {Schlom}},\ and\ \bibinfo {author} {\bibfnamefont {K.~M.}\ \bibnamefont {Shen}},\ }\bibfield  {title} {\bibinfo {title} {Incoherent cooper pairing and pseudogap behavior in single-layer $\mathrm{FeSe}/\mathrm{SrTi}{\mathrm{o}}_{3}$},\ }\href {https://doi.org/10.1103/PhysRevX.11.021054} {\bibfield  {journal} {\bibinfo  {journal} {Phys. Rev. X}\ }\textbf {\bibinfo {volume} {11}},\ \bibinfo {pages} {021054} (\bibinfo {year} {2021})}\BibitemShut {NoStop}%
\bibitem [{\citenamefont {Rasmussen}\ and\ \citenamefont {Williams}(2006)}]{Rasmussen2006Gaussian}%
  \BibitemOpen
  \bibfield  {author} {\bibinfo {author} {\bibfnamefont {C.~E.}\ \bibnamefont {Rasmussen}}\ and\ \bibinfo {author} {\bibfnamefont {C.~K.~I.}\ \bibnamefont {Williams}},\ }\href@noop {} {\emph {\bibinfo {title} {Gaussian Processes for Machine Learning}}}\ (\bibinfo  {publisher} {The MIT Press},\ \bibinfo {year} {2006})\BibitemShut {NoStop}%
\bibitem [{\citenamefont {Gardner}\ \emph {et~al.}(2021)\citenamefont {Gardner}, \citenamefont {Pleiss}, \citenamefont {Bindel}, \citenamefont {Weinberger},\ and\ \citenamefont {Wilson}}]{gardner2021gpytorchblackboxmatrixmatrixgaussian}%
  \BibitemOpen
  \bibfield  {author} {\bibinfo {author} {\bibfnamefont {J.~R.}\ \bibnamefont {Gardner}}, \bibinfo {author} {\bibfnamefont {G.}~\bibnamefont {Pleiss}}, \bibinfo {author} {\bibfnamefont {D.}~\bibnamefont {Bindel}}, \bibinfo {author} {\bibfnamefont {K.~Q.}\ \bibnamefont {Weinberger}},\ and\ \bibinfo {author} {\bibfnamefont {A.~G.}\ \bibnamefont {Wilson}},\ }\href {https://arxiv.org/abs/1809.11165} {\bibinfo {title} {Gpytorch: Blackbox matrix-matrix gaussian process inference with gpu acceleration}} (\bibinfo {year} {2021}),\ \Eprint {https://arxiv.org/abs/1809.11165} {arXiv:1809.11165 [cs.LG]} \BibitemShut {NoStop}%
\bibitem [{\citenamefont {Greenhill}\ \emph {et~al.}(2020)\citenamefont {Greenhill}, \citenamefont {Rana}, \citenamefont {Gupta}, \citenamefont {Vellanki},\ and\ \citenamefont {Venkatesh}}]{Greenhill2020}%
  \BibitemOpen
  \bibfield  {author} {\bibinfo {author} {\bibfnamefont {S.}~\bibnamefont {Greenhill}}, \bibinfo {author} {\bibfnamefont {S.}~\bibnamefont {Rana}}, \bibinfo {author} {\bibfnamefont {S.}~\bibnamefont {Gupta}}, \bibinfo {author} {\bibfnamefont {P.}~\bibnamefont {Vellanki}},\ and\ \bibinfo {author} {\bibfnamefont {S.}~\bibnamefont {Venkatesh}},\ }\bibfield  {title} {\bibinfo {title} {Bayesian optimization for adaptive experimental design: A review},\ }\href@noop {} {\bibfield  {journal} {\bibinfo  {journal} {IEEE Access}\ }\textbf {\bibinfo {volume} {8}},\ \bibinfo {pages} {13937} (\bibinfo {year} {2020})}\BibitemShut {NoStop}%
\bibitem [{\citenamefont {Kingma}\ and\ \citenamefont {Ba}(2017)}]{kingma2017adammethodstochasticoptimization}%
  \BibitemOpen
  \bibfield  {author} {\bibinfo {author} {\bibfnamefont {D.~P.}\ \bibnamefont {Kingma}}\ and\ \bibinfo {author} {\bibfnamefont {J.}~\bibnamefont {Ba}},\ }\href {https://arxiv.org/abs/1412.6980} {\bibinfo {title} {Adam: A method for stochastic optimization}} (\bibinfo {year} {2017}),\ \Eprint {https://arxiv.org/abs/1412.6980} {arXiv:1412.6980 [cs.LG]} \BibitemShut {NoStop}%
\bibitem [{\citenamefont {Noack}\ \emph {et~al.}(2019)\citenamefont {Noack}, \citenamefont {Yager}, \citenamefont {Fukuto}, \citenamefont {Doerk}, \citenamefont {Li},\ and\ \citenamefont {Sethian}}]{noack2019kriging}%
  \BibitemOpen
  \bibfield  {author} {\bibinfo {author} {\bibfnamefont {M.~M.}\ \bibnamefont {Noack}}, \bibinfo {author} {\bibfnamefont {K.~G.}\ \bibnamefont {Yager}}, \bibinfo {author} {\bibfnamefont {M.}~\bibnamefont {Fukuto}}, \bibinfo {author} {\bibfnamefont {G.~S.}\ \bibnamefont {Doerk}}, \bibinfo {author} {\bibfnamefont {R.}~\bibnamefont {Li}},\ and\ \bibinfo {author} {\bibfnamefont {J.~A.}\ \bibnamefont {Sethian}},\ }\bibfield  {title} {\bibinfo {title} {A kriging-based approach to autonomous experimentation with applications to x-ray scattering},\ }\href@noop {} {\bibfield  {journal} {\bibinfo  {journal} {Scientific Reports}\ }\textbf {\bibinfo {volume} {9}},\ \bibinfo {pages} {11809} (\bibinfo {year} {2019})}\BibitemShut {NoStop}%
\bibitem [{\citenamefont {Noack}\ \emph {et~al.}(2020)\citenamefont {Noack}, \citenamefont {Doerk}, \citenamefont {Li}, \citenamefont {Streit}, \citenamefont {Vaia}, \citenamefont {Yager},\ and\ \citenamefont {Fukuto}}]{noack2020autonomous}%
  \BibitemOpen
  \bibfield  {author} {\bibinfo {author} {\bibfnamefont {M.~M.}\ \bibnamefont {Noack}}, \bibinfo {author} {\bibfnamefont {G.~S.}\ \bibnamefont {Doerk}}, \bibinfo {author} {\bibfnamefont {R.}~\bibnamefont {Li}}, \bibinfo {author} {\bibfnamefont {J.~K.}\ \bibnamefont {Streit}}, \bibinfo {author} {\bibfnamefont {R.~A.}\ \bibnamefont {Vaia}}, \bibinfo {author} {\bibfnamefont {K.~G.}\ \bibnamefont {Yager}},\ and\ \bibinfo {author} {\bibfnamefont {M.}~\bibnamefont {Fukuto}},\ }\bibfield  {title} {\bibinfo {title} {Autonomous materials discovery driven by gaussian process regression with inhomogeneous measurement noise and anisotropic kernels},\ }\href@noop {} {\bibfield  {journal} {\bibinfo  {journal} {Scientific Reports}\ }\textbf {\bibinfo {volume} {10}},\ \bibinfo {pages} {17663} (\bibinfo {year} {2020})}\BibitemShut {NoStop}%
\bibitem [{\citenamefont {Mnih}\ \emph {et~al.}(2015)\citenamefont {Mnih}, \citenamefont {Kavukcuoglu}, \citenamefont {Silver}, \citenamefont {Rusu}, \citenamefont {Veness}, \citenamefont {Bellemare}, \citenamefont {Graves}, \citenamefont {Riedmiller}, \citenamefont {Fidjeland}, \citenamefont {Ostrovski},\ and\ \citenamefont {Petersen}}]{Mnih2015Human}%
  \BibitemOpen
  \bibfield  {author} {\bibinfo {author} {\bibfnamefont {V.}~\bibnamefont {Mnih}}, \bibinfo {author} {\bibfnamefont {K.}~\bibnamefont {Kavukcuoglu}}, \bibinfo {author} {\bibfnamefont {D.}~\bibnamefont {Silver}}, \bibinfo {author} {\bibfnamefont {A.~A.}\ \bibnamefont {Rusu}}, \bibinfo {author} {\bibfnamefont {J.}~\bibnamefont {Veness}}, \bibinfo {author} {\bibfnamefont {M.~G.}\ \bibnamefont {Bellemare}}, \bibinfo {author} {\bibfnamefont {A.}~\bibnamefont {Graves}}, \bibinfo {author} {\bibfnamefont {M.}~\bibnamefont {Riedmiller}}, \bibinfo {author} {\bibfnamefont {A.~K.}\ \bibnamefont {Fidjeland}}, \bibinfo {author} {\bibfnamefont {G.}~\bibnamefont {Ostrovski}},\ and\ \bibinfo {author} {\bibfnamefont {S.}~\bibnamefont {Petersen}},\ }\bibfield  {title} {\bibinfo {title} {Human-level control through deep reinforcement learning},\ }\href@noop {} {\bibfield  {journal} {\bibinfo  {journal} {Nature}\ }\textbf {\bibinfo {volume} {518}},\ \bibinfo {pages} {529} (\bibinfo {year} {2015})}\BibitemShut {NoStop}%
\bibitem [{\citenamefont {Kamthe}\ and\ \citenamefont {Deisenroth}(2018)}]{kamthe2018data}%
  \BibitemOpen
  \bibfield  {author} {\bibinfo {author} {\bibfnamefont {S.}~\bibnamefont {Kamthe}}\ and\ \bibinfo {author} {\bibfnamefont {M.}~\bibnamefont {Deisenroth}},\ }\bibfield  {title} {\bibinfo {title} {Data-efficient reinforcement learning with probabilistic model predictive control},\ }in\ \href@noop {} {\emph {\bibinfo {booktitle} {International Conference on Artificial Intelligence and Statistics}}}\ (\bibinfo {organization} {PMLR},\ \bibinfo {year} {2018})\ pp.\ \bibinfo {pages} {1701--1710}\BibitemShut {NoStop}%
\end{thebibliography}%
\end{document}